\documentclass[a4paper,12pt]{article}

\usepackage{jheppub} 

\usepackage[T1]{fontenc} 
\usepackage{bm,latexsym,amsmath,amssymb,amsfonts,mathtools,mathrsfs}
\usepackage{comment}
\usepackage{ulem}
\usepackage{color}

\usepackage{bm}
\usepackage{hyperref}
\usepackage{framed}
\usepackage{subfigure}
\usepackage{hyperref}   
\usepackage{graphicx}

\expandafter\let\csname equation*\endcsname\relax 
\expandafter\let\csname endequation*\endcsname\relax 
\usepackage{amsmath}
\usepackage{color}

\usepackage{hyperref}
\hypersetup{
    colorlinks=true,
    linkcolor=blue,
    filecolor=magenta,
    urlcolor=blue,
}
\usepackage{float}
\usepackage{ifthen}
\usepackage{xspace}
\usepackage{relsize}
\usepackage{yhmath}
\usepackage[left]{lineno}

\newcommand{\beqa}{\begin{eqnarray}}
\newcommand{\eeqa}{\end{eqnarray}}
\newcommand{\ba}{\begin{array}}
\newcommand{\ea}{\end{array}}

\def\d{\text{d}}
\def\dis{\displaystyle}
\def\({\left(}
\def\){\right)}
\def\ro{r_{\!o}^{}}
\def\CC{\Lambda}
\def\rc{r_{\text{cri}}^{}}
\def\rb{\bar{r}}

\begin{document}

\title{Geometrization of light bending and its application to SdS$_w$ spacetime} 

\author[*]{Zhen Zhang}
\emailAdd{zhangzhen@ihep.ac.cn}
\affiliation[*]{Key Laboratory of Particle Astrophysics, Institute of High Energy Physics, Chinese Academy of Sciences, 19B Yuquan Road, Beijing 100049, People's Republic of China}

\date{\today}

\abstract{
The mysterious dark energy remains one of the greatest puzzles of modern science. Current detections for it are mostly indirect. The spacetime effects of dark energy can be locally described by the SdS$_w$ metric. Understanding these local effects exactly is an essential step towards the direct probe of dark energy. From first principles, we prove that dark energy can exert a repulsive dark force on astrophysical scales, different from the Newtonian attraction of both visible and dark matter. One way of measuring local effects of dark energy is through the gravitational deflection of light. We geometrize the bending of light in any curved static spacetime. First of all, we define a generalized deflection angle, referred to as the Gaussian deflection angle, in a mathematically strict and conceptually clean way. Basing on the Gauss-Bonnet theorem, we then prove that the Gaussian deflection angle is equivalent to the surface integral of the Gaussian curvature over a chosen lensing patch. As an application of the geometrization, we study the problem of whether dark energy affects the bending of light and provide a strict solution to this problem in the SdS$_w$ spacetime. According to this solution, we propose a method to overcome the difficulty of measuring local dark energy effects. Exactly speaking, we find that the lensing effect of dark energy can be enhanced by 14 orders of magnitude when properly choosing the lensing patch in certain cases. It means that we can probe the existence and nature of dark energy directly in our Solar System. This points to an exciting direction to help unraveling the great mystery of dark energy.
}

\maketitle

\section{Introduction}

Dark energy composes about 69\% of the total energy density of the present universe. 
It is almost 15 times larger than all the visible matter we see in our universe.
But so far the existing evidences of dark energy are mostly indirect \cite{DEexp}. 
The mystery of dark energy poses a great challenge to modern science.  
Since the cosmological constant $\Lambda$ was introduced by Einstein into his field equation of general relativity (GR),
it has long been viewed as representing dark energy. 
Up to now, various models of dark energy (such as scalar fields) have been proposed to interpret the current astronomy observations \cite{DEmodel}.

NASA has already proposed a developed mission on the gravity observation and dark energy detection explorer in the Solar System (GODDESS)
by flying a constellation of long-baseline atom-interferometer gravity gradiometers 
and measuring the trace of the force field gradient tensor in the Solar System \cite{Nan18}. 
It will possibly be able to isolate the new force field signal from overwhelmingly stronger gravity effects and achieve a direct detection of dark energy as a scalar field \cite{Nan18}.

Perhaps, dark energy has already been detected through non-gravitational effects \cite{Vagnozzi21}. 
The coupling of dark energy to photons leads to its production in the strong magnetic field of the solar tachocline via a mechanism analogous to the Primakoff process for axions. 
This allows for detectable signals on Earth. In fact, the electron recoil excess recently reported by the XENON1T \cite{XENON-1T} (deep underground at the INFN laboratory, Italy) collaboration has been explained by chameleon-screened dark energy, indicating the first direct detection of dark energy \cite{Vagnozzi21}. 
Future detectors such as XENONnT \cite{XENONnT}, PandaX-4T \cite{Zhang4T}, and LUX-ZEPLIN \cite{LUXZEPLIN} are on the way.

Early in 2017, He and Zhang proposed to probe dark energy directly through gravitational effects \cite{HZ2017}.
In general, dark energy is characterized by an equation-of-state parameter $w$, which may evolve with the cosmological redshift $z$, namely $w=w\(z\)$ \cite{DEmodel}.
Different models of dark energy can be described by the parameter $w$ when it takes the corresponding model values,  
such as the cosmological constant model ($\,w=-1\,$), the quintessence model ($-1<w <-\frac{1}{3}\,$), and 
the phantom model ($\,w<-1$) \cite{DEmodel,HZ2017}. However, $w$ can be treated as a constant on astrophysical scales, like our Solar System \cite{HZ2017}.
Based on this, it was proved model-~and~state-independently in He~\&~Zhang~$\(2017\)$ that the local repulsion from dark energy can be described well by the SdS$_{w}$ metric \cite{HZ2017},
\beqa
\label{eq:dSS2}
\nonumber
\d S_{w}^2 \,=\,\!&-& \bigg[1\!-2\,\frac{\,M\,}{r}-2\left( \frac{\,\ro\,}{r} \right)^{\!3w+1}\bigg]\,\d t^{2}\\[2mm]
&+&\frac{1}{\, \bigg[1-2\,\frac{\,M\,}{r}-2\( \frac{\,\ro\,}{r} \)^{\!3w+1}\bigg] \,}\,\d r^2\!+ r^{2}\left(\d \theta^2+\sin^{2}\theta\,\d\phi^2\right),
\eeqa
where the mass parameter $M$ is determined by both visible and dark matter of the gravitational system, 
$\,r_o^{}\,$ is a model-parameter characterizing the size of the present universe, and $(t,~r,~\theta,~\phi)$ are the spherical (spacetime) coordinates.
Here we adopt the geometrized unit system ($G=c=1$).  
At large $r$, the SdS$_{w}$ spacetime can be conformally and isometrically embedded into the Friedmann-Robertson-Walker spacetime \cite{HZ2017}.
Besides, the scale factor $\,r_o^{}\,$ can be given in each dark energy model when comparing with the cosmological data; 
it takes the value $\,\ro = \sqrt{6/\CC}\,$ for the cosmological constant model only and not for general dark energy models \cite{HZ2017}.
By definition, the pressure $p_{i}$, the energy density $\rho_{i}$, and the equation-of-state parameter $w$ are connected by
\beqa
\ba{rcl}
\dis
w &=& \sum\limits_{i}\,p_{i}\,\bigg/\sum\limits_{i}\,\rho_{i} <-\frac{1}{3},
\ea
\eeqa
where $i$ represents the $i$th contribution described by a different dark energy model \cite{DEmodel}. Note that there is only one $w$-term responsible for all these dark energy contributions in the SdS$_{w}$ metric. When $w=-1$, the metric \eqref{eq:dSS2} reduces to the well-known Schwarzschild-de Sitter (SdS) one.

For the SdS case, various kinds of studies on $\Lambda$ have been carried out previously. 
In the early 1980s, Islam \cite{Islam83} showed that the light orbital equation is independent of $\Lambda$ in the SdS spacetime. 
Since then it was generally believed that dark energy plays no role on the bending of light. By 2007, Rindler~and~Ishak 
had concluded that $\Lambda$ contributes to the deflection of light after considering the measurements done by observers \cite{RI2007} . 
However, the conclusion has led to a debate of more than ten years \cite{Ishak2008,Arakida12,Italy,Sereno08,Bhadra10,other1,Ishak2010,Ishak-Rev}. 
The difficulty of understanding the influence of dark energy on light bending is due to the fact that the SdS spacetime is not flat at spatial infinity. 
To overcome this difficulty, the Gauss-Bonnet theorem have been applied to a special case of the static spacetime of spherical symmetry in the literature \cite{Gibbons08,Ishihara16,Ishihara17,Arakida18}. 
However, the difficulty has still not been resolved completely in the SdS spacetime; some concepts and definitions remain to be clarified \cite{Arakida18}.
Ten years later in 2017, He and Zhang \cite{HZ2017} extended the debate to the general SdS$_{w}$ spacetime.

In this work, we attempt to geometrize the bending of light using the Gauss-Bonnet theorem and propose a method to overcome the difficulty of measuring local dark energy effects.
In order to illustrate our basic ideas in a clear way, we just focus on the ideal SdS$_{w}$ case in which a point-like mass is surrounded by dark energy with a generic state parameter $w<-\frac{1}{3}$.
In fact, when $w\geq-\frac{1}{3}$, the metric \eqref{eq:dSS2} is still a solution to the Einstein equation,  
which can be verified straightforwardly by following the derivations presented in He~\&~Zhang~$\(2017\)$ \cite{HZ2017}. 
So the SdS$_{w}$ metric is also applicable to the case of $w\geq-\frac{1}{3}$.
However, in this case, the metric becomes asymptotically flat, and thus it is quite trivial to understand the role of the $w$-term in the bending of light.

This work is organized as follows. In section~\ref{sec:2}, we introduce the projection tensors. In section~\ref{sec:3}, 
we analyze the dark energy effect on massive particles, and demonstrate that it acts as a kind of gravitational force in the SdS$_{w}$ spacetime. In section~\ref{sec:4}, by presenting new techniques, 
we perform the geometrization for the bending of light, and extend the concept of light deflection to any curved static spacetime. 
In section~\ref{sec:5}, we give a strict solution to the problem of the influence of dark energy on light bending in the SdS$_{w}$ spacetime, and propose a method to overcome the difficulty of measuring the local dark energy effect. We conclude in section~\ref{sec:6}. In appendices \ref{app:A}, \ref{app:B}, \ref{app:C}, \ref{app:D} and \ref{app:E}, we present additional results and explanations of our theories that clarify and support the results in the main text.

\vspace*{3mm}
\section{\hspace*{+0.0mm}Projection tensors}
\label{sec:2}
\vspace*{1mm}

For any observer, let $U^{a}$ be their four-velocity with $U^{a}U_{a}=-1$, where $a$ marks the abstract index notation. For convenience, we will use the notation through this work, which was widely used in Wald's book\,\cite{Wald84}.  
Let $h_{ab}=g_{ab}+U_{a}U_{b}$ and $\pi_{ab}=-\,U_{a}U_{b}$, where $g_{ab}$ is the metric tensor used to describe the entire spacetime. Then, we have 
\beqa
\ba{rcl}
\dis
\nonumber
h_{ab}\,h^{b}_{\,\,\,c} &=& h_{ac},~~ h_{ab}\,\pi^{b}_{\,\,\,c} = 0,~~\pi_{ab}\,\pi^{b}_{\,\,\,c} = \pi_{ac}.
\ea
\eeqa
Indeed, $h_{ab}$ and $\pi_{ab}$ are space-like and time-like projection tensors, respectively. 
And we have 
\beqa
\ba{rcl}
\dis
\nonumber
h^{a}_{\,\,\,b}\,U^{b}=0,\,\,h_{ab}\,U^{b}=0,\,\,\pi^{a}_{\,\,\,b}\,U^{b}=U^{a},~\pi_{ab}\,U^{b}=U_{a}.
\ea
\eeqa
For a given four-vector $V^{a}$, let $\,V^{\,\,\,a}_{\parallel}\,$ and $\,V^{\,\,\,a}_{\perp}\,$ be the components which are parallel and perpendicular to $\,U^{a}$, respectively. Then we obtain
\beqa
\label{eq:Vperp}
\ba{rcl}
\dis
V^{\,\,\,a}_{\perp} &=& h^{a}_{\,\,\,b}\,V^{b},~~V^{\,\,\,a}_{\parallel} = \pi^{a}_{\,\,\,b}\,V^{b},\\[3mm]
V_{\perp \,a} &=& h_{ab}\,V^{b},~~ V_{\parallel \,a} =\pi_{ab}\,V^{b},
\ea
\eeqa
where $V_{a}$ is the corresponding dual vector. Usually, one has $V_{a}=g_{ab}\,V^{b}$. Clearly, $\,h^{a}_{\,\,\,b}\,$ projects the four-vector onto the local space of the observer. 
It is noteworthy that $h_{ab}$ is actually the metric tensor of some subspace of the entire spacetime.
Denote $\,\Xi\,$ as this subspace, and parametrize it by a set of coordinates $z^{\mu}$, $\mu=0, 1, 2, 4$. The tensors $\,g_{ab}\,$ and $\,h_{ab}\,$ can be expressed in general as 
\beqa
\label{eq:gab}
g_{ab} &=& g_{\mu\nu}\,\(\d z^{\mu}\)_{a}\,\(\d z^{\nu}\)_{b},\\[1mm]
\pi_{ab} &=& \pi_{\mu\nu}\,\(\d z^{\mu}\)_{a}\,\(\d z^{\nu}\)_{b},\\[1mm]
\label{eq:hab}
h_{ab} &=& h_{\mu\nu}\,\(\d z^{\mu}\)_{a}\,\(\d z^{\nu}\)_{b},
\eeqa
respectively, where $\pi_{\mu\nu}= -\,U_{\mu}\,U_{\nu}$ and $h_{\mu\nu}=g_{\mu\nu}+U_{\mu}\,U_{\nu}$.  Thus, $h_{ab}$ can be induced directly from $g_{ab}$, namely $h_{ab}=g_{ab}\big|_{\Xi}$.
Accordingly, it can be described well in the same coordinate system as $\,g_{ab}$. 
In a similar way, $V^{a}$ and $V_{a}$ can be written as 
\beqa
\label{eq:Va}
V^{a} = V^{\mu}\,\,\partial_{\mu}^{\,\,\,a},~~~~\textit{V}_{\textit{a}} = \textit{V}_{\mu}\,\(\d \textit{z}^{\mu}\)_{\textit{a}},
\eeqa
with $\partial_{\mu}^{\,\,\,a}=\(\partial/\partial z^{\mu}\)^{a}$. 
Note that, for $h_{ab}$, all its Lorentz indices are contracted. 
So the subspace described by $h_{ab}$ is actually a physical space in which measurements can be made by the observer.
Let $V^{\mu}_{\perp}=h^{\mu}_{\,\,\,\nu} \,V^{\nu}$, and $V_{\perp \,\mu}=h_{\mu\nu} \,V^{\nu}$. 
Then we have
\beqa
\label{eq:Vperp1}
V_{\perp}^{\,\,\,a} &=& V^{\mu}_{\perp}\,\,\partial_{\mu}^{\,\,\,a},~~~V_{\perp \,a} = V_{\perp \,\mu}\,\(\d z^{\mu}\)_{a}.
\eeqa
Clearly, any four-vector $V^{a}$ can be projected into the local space of the observer. 
For the projected $\,V_{\perp}^{\,\,\,a}\,$ by $\,h^{a}_{\,\,\,b}\,$ and $\,V_{\perp \,a}\,$ by $\,h_{ab}$, they are both three-vectors with Lorentz indices contracted.
From the viewpoint of physics, they are both physical quantities and can be measured by the observer. 
So those quantities derived directly from them, such as the intersection angle given by equation\,\eqref{eq:thetaM} of appendix \ref{app:A}, 
are all independent of the choice of coordinates and thus physically measurable, especially in our notations and conventions.

\vspace*{3mm}
\section{\hspace*{+0.0mm}The Newtonian analogy}
\label{sec:del}
\label{sec:3}
\vspace*{1mm}

In GR, the Einstein equation determining the motion of matter allows a Newtonian interpretation, and thus a degree of intuitive understanding, which is often unavailable from the formalism alone.
Let us consider the Newtonian motion of a massive test-particle in the gravitational field including both contributions of the mass source and dark energy. 
In Newtonian gravity, the test-particle experiences an acceleration in a gravitational field, and thus can be used to help us 
obtain the exact form of the attractive Newtonian force of the mass source (including both visible and dark matter). 
The force induced by dark energy can be interpreted in a quite similar way to the attractive Newtonian force. 
In principle, we can obtain the exact form of the \textit{total gravitational force} including both the Newtonian attraction and the dark energy contribution from GR, starting from the metric tensor.

To begin with, we define the Newtonian limit by three requirements: the gravitational field is weak, and it is static 
as well as the particles are moving very slowly compared to the speed of light \cite{Carroll14}. 
Next, let us see how to introduce the total gravitational force including contributions from both the mass source and dark energy. 
Firstly, its form should be intrinsically determined by the metric tensor $g_{ab}$. 
Secondly, it should allow to reproduce the conventional results of Newtonian gravity in the Newtonian limit. 
Thirdly, it should not include any information about the motion of the observer or the test-particle so that the total gravitational force is intrinsic to the spacetime itself. 
When taking all these requirements into consideration, 
we will find that only as an instantaneous observer who is static relative to the gravitational field, 
one can feel the gravitational field directly and thus obtain the exact form of the total gravitational force correctly in the Newtonian limit.

In theories of stationary spacetime, for any observer, their four-acceleration is usually nonzero and given by $\hat{A}^{a}=U^{b}\nabla_{b}U^{a}$, 
where $\nabla_{b}$ is the covariant derivative operator associated with the metric tensor $g_{ab}$ \cite{Wald84}. 
Let $V^{a}$ be the four-velocity of a test-particle moving on a geodesic and passing through the local space of the observer, then the observed three-acceleration by this observer
would be $\hat{a}^{a}=-\hat{A}^{a}+\frac{2}{(V^{c}U_{c})^2}\,(V^{b}\hat{A}_{b})\,V_{\perp}^{\,\,\,a}$.
Setting $V^{a}=U^{a}$, we have $\hat{a}^{a}=-\hat{A}^{a}$.
In this case, the three-acceleration $\hat{a}^{a}$ is measured in the rest frame of the test-particle, usually named as the \textit{proper acceleration} \cite{Rindler06}. 
However, the proper acceleration is still observer dependent.
If the instantaneous observer is static relative to the gravitational field, 
the measured proper acceleration $\hat{a}^{a}$ by this observer is just the gravitational three-force $\vec{g}$ on per unit mass of the particle, especially in the Newtonian limit \cite{Wald84,Carroll14,Rindler06}.
The specific $\vec{g}$-form obtained directly from this proper acceleration will be referred to as the \textit{proper form} of the gravitational force. 
Clearly, the measured proper acceleration $\hat{a}^{a}$ by the observer is uniquely determined by the metric tensor,
and, therefore, so is the proper form of the gravitational three-force $\vec{g}$.

In the case of some small physical separation, $\delta^{a}$, between the test-particle and observer, 
we have the three-acceleration, $\hat{a}^{a}=- (1+\delta^{b}\nabla_{b})\, \hat{A}^{a} +R^{a}_{\,\,\,bcd}\,\,U^{b}U^{c}\delta^{d}$, where $R^{a}_{\,\,\,bcd}$ is the Riemann tensor of $\(1, 3\)$-type. 
Thus, the separation in this case will lead to some additional $\delta$-terms. 
However, in GR, the measurements have to be made by the instantaneous observer at the point of the test-particle or a sufficiently small region around the test-particle. 
So $\delta^{a}$ is very small in an actual measurement, and thus these $\delta$-terms can be ignored directly. 
Alternatively, the proper acceleration can be obtained by making $\delta$-corrections to the measured acceleration by the observer located at a small separation of $\delta^{a}$ to the test-particle.

As shown in equation\,\eqref{eq:dSS2}, the SdS$_{w}$ metric is static. So there exists a time-like Killing vector $\cal{K}^{\textit{a}}\!=\!(\partial/\partial\,\!\textit{t})^{\textit{a}}$. Let $\chi\!=\!\sqrt{-\cal{K}_{\textit{a}}\cal{K}^{\textit{a}}}$. Here the range that we are interested in is $1\!-\!2\frac{\,M\,}{r}\!-\!2\left(\! \frac{\,\ro\,}{r} \!\right)^{\!3w\!+\!1}\!>\!0$. Then, $\chi\!=\!\sqrt{-g_{00}}=\sqrt{1\!-\!2\frac{\,M\,}{r}\!-\!2\left(\! \frac{\,\ro\,}{r} \!\right)^{\!3w\!+\!1}}$.
By definition, we have $\hat{A}^{a}=\nabla^{a}\rm{ln} \,\chi$ \cite{Wald84}. 
Correspondingly, $\hat{A}_{a}=\nabla_{a}\rm{ln} \,\chi=\({\it d}\,\rm{ln} \,\chi\)_{\it{a}}=\frac{1}{\sqrt{-\it{g_{{\rm 00}}}}}\,\frac{\partial\,\!\sqrt{-\it{g_{{\rm 00}}}}}{\partial\,\!{\it r}}\,\(\d\,\!\it{r}\)_{\it{a}}$, 
where $\({\it d}\,\rm{ln} \,\chi\)_{a}$ is the total differential of $\rm{ln}\,\chi$.
We therefore obtain the proper form of the three-acceleration of the massive test-particle as follows, 
\beqa
\label{eq:4Aa}
\nonumber
\hat{a}^{a}\!=-\hat{A}^{a}&=&-g^{ab}\,\hat{A}_{b}=-\(g^{\mu\nu}\,\partial_{\mu}^{\,\,\,a}\,\partial_{\nu}^{\,\,\,b}\)\,\left[\frac{\,\,1}{\sqrt{-g_{00}}}\,\frac{\partial\!\,\sqrt{-g_{00}}}{\partial\,\!r}\,\(\d\,\!r\)_{b}\right]\\[3mm]
\nonumber
&=&-\frac{\,\,g^{rr}}{\sqrt{-g_{00}}}\,\frac{\partial\!\,\sqrt{-g_{00}}}{\partial\,\!r}\,\(\frac{\partial}{\partial\,\!r}\)^{\!a}\\[3mm]
\nonumber
&=&-\frac{1}{r}\left[\frac{\,M\,}{r}\!+\!(3w\!+\!1)\left(\! \frac{\,\ro\,}{r} \!\right)^{\!3w\!+\!1}\right]\,\(\frac{\partial}{\partial\,\!r}\)^{\!a}\\[3mm]
&=&-\frac{1}{r}\frac{1}{\sqrt{1\!-\!2\frac{\,M\,}{r}\!-\!2\left(\! \frac{\,\ro\,}{r} \!\right)^{\!3w\!+\!1}}}\,\left[\frac{\,M\,}{r}\!+\!(3w\!+\!1)\left(\! \frac{\,\ro\,}{r} \!\right)^{\!3w\!+\!1}\right]\textit{e}_{r}^{\,a}, 
\hspace*{10mm}
\eeqa
where  $\textit{e}_{r}^{\,a}=\frac{1}{\sqrt{g_{rr}}}\(\frac{\partial}{\partial\,\!r}\)^{\!a}$ is the unit vector along the radial direction. 
In our notations, we have $g^{ab}=g^{\mu\nu}\,\partial_{\mu}^{\,\,\,a}\,\partial_{\nu}^{\,\,\,b}$ and $\(\d z^{\nu}\)_{b}\,\partial_{\mu}^{\,\,\,b}\,=\delta^{\,\,\nu}_{\mu}$, 
which have been used in the first and second lines, respectively.
From equation\eqref{eq:4Aa}, we see that besides the attractive Newtonian force, there exits a {\it dark force} (generated by dark energy) on the massive test-particle.
Evidently, the dark force and the Newtonian force are on equal footing.
The observational data requires $\,3w+2 < 0\,$ at $\,6\,\sigma$\, level\,\cite{2015BAO}. 
Thus, the $w-$term leads to a repulsive dark force. 
For the special case of the cosmological constant $\CC$, we have $\,\ro = \sqrt{6/\CC}\,$.
In this case, we can rewrite equation\,\eqref{eq:4Aa} by setting $w=-1$ in a more familiar form
\beqa
\label{eq:g3force}
\nonumber
\vec{g}= -\frac{1}{r}\frac{1}{\sqrt{1\!-\!2\frac{\,M\,}{r}\!-2\frac{\CC}{\,6\,}r^2}}\(\frac{\,M\,}{\,\,r\,}\!-\!\frac{\CC}{3}r^2\)\,\boldsymbol{\hat{r}},
\hspace*{10mm}
\eeqa
with $|\boldsymbol{\hat{r}}|=1$. 
Under the weak field approximation, one gets ~$\vec{g}\,\simeq-\(\frac{\,M\,}{\,\,r^2\,}\!-\!\frac{\CC}{3}r\)\,\boldsymbol{\hat{r}}$, 
which agrees with the dark force shown by Ho and Hsu in \,\cite{Ho15}. 
Now we successfully obtain the proper form of the total gravitational force via equation\,\eqref{eq:4Aa}, independently of specific dark energy models.

As seen from the static observer with respect to the gravitational field, 
the strength of the three-force $\,\vec{g}\,$ on the test-particle, also known as the gravitational field strength, 
equals the magnitude of the three-acceleration \eqref{eq:4Aa}, namely $g=\left|\vec{g}\right|=\left|\hat{a}^{a}\right|$. 
Exactly, the gravitational field strength $g$ can be written as 
\beqa
\label{eq:gforce}
\nonumber
g \,&=&\sqrt{\hat{a}^{a}\,\hat{a}_{a}}=\sqrt{\hat{A}^{a}\,\hat{A}_{a}}\\[3mm]
\nonumber
&=&\sqrt{\left[\frac{\,\,g^{rr}}{\sqrt{-g_{00}}}\,\frac{\partial\!\,\sqrt{-g_{00}}}{\partial\,\!r}\,\(\frac{\partial}{\partial\,\!r}\)^{\!a}\right] \left[\frac{1}{\sqrt{-g_{00}}}\,\frac{\partial\,\!\sqrt{-g_{00}}}{\partial\,\!r}\,\(\d\,\!r\)_{a}\right]}\\[3mm]
\nonumber
&=&\!\bigg|\frac{1}{\sqrt{\(-g_{00}\)g_{rr}}\,\,}\,\(\frac{\partial\,\!\sqrt{-g_{00}}}{\partial\,\!r}\)\!\bigg|\\[3mm]
&=&\frac{1}{\,r\,}\frac{\,1\,}{\sqrt{1\!-\!2\frac{\,M\,}{r}\!-\!2\left(\! \frac{\,\ro\,}{r} \!\right)^{\!3w\!+\!1}}}\,\bigg|\!\frac{\,M\,}{r}\!+\!(3w\!+\!1)\left(\! \frac{\,\ro\,}{r} \!\right)^{\!3w\!+\!1}\!\bigg|,
\hspace*{10mm}
\eeqa
where we have used $\(\d\,\!r\)_{a}\(\frac{\partial}{\partial\,\!r}\)^{\!a}=1$ in the third line.
When taking $\ro\to\infty$, one has 
\beqa
\label{eq:CCforce}
\nonumber
g\simeq\frac{\,M\,}{r^2}\frac{1}{\sqrt{1\!-\!2\frac{\,M\,}{r}}},
\hspace*{10mm}
\eeqa
which is just the Rindler's form of the field strength shown by equation\,$\(11.15\)$ in \,\cite{Rindler06} for the special case of the Schwarzschild spacetime. 
In this case, we see that, for large $r$, the field strength recovers the Newtonian inverse-square law in the Newtonian limit, namely $g\!\simeq\!\frac{\,\,M\,\,\,}{\,r^2\,}$. 
It is important to mention that the SdS$_{w}$ metric is true not only for the cosmological constant but also for other possible forms of dark energy (such as a single effective scalar field). 
In general, there is no explicit cosmological constant $\CC$ involved in the SdS$_{w}$ metric; 
that is, the parameter $\ro$ can not be expressed in terms of $\CC$ for general dark energy models.
Additionally, we do not have the explicit $\ro$ value for each dark energy model at present. Hence, the Newtonian analogue cannot be achieved directly from arbitrary state parameter $w$. 
One has to first set it to $w=-1$, i.e., the cosmological constant case. Thereafter one should set $\CC=0$ (to achieve $\ro=\infty$), and finally for $r\gg\,M$, one recovers the Newtonian analogy. 
Anyway, we have shown that the three-force $\vec{g}$ (or, equivalently, $\hat{a}^{a}$) given by equation\eqref{eq:4Aa} satisfies all the requirements mentioned earlier, 
and we therefore successfully obtain the proper form of the total gravitational force including dark energy.

The dark force acts on astrophysical scales (such as the Solar-System) that can be tens of orders of magnitudes lower than the cosmological scale.
Requiring $\vec{g}=0$ or $g=0$, we derive the critical radius $\rc$ without any further approximations, 
\beqa
\rc \,=\, \ro\(\!\frac{M}{\,\left|3w\!+\!1\right|\,\ro\,}\!\)^{\!\!-\frac{1}{3w}}
\label{Arc}
\label{eq:Arc}
\eeqa
where the repulsive dark force can balance the attractive Newtonian force. 
This coincides with Ho\,\,\&\,\,Hsu~$\(2015\)$ for the cosmological constant model\,\cite{Ho15} and with He\,\,\&\,\,Zhang~$\(2017\)$ for general dark energy models \cite{HZ2017}. 
However, the post-Newtonian approximation adopted in He~\&~Zhang~$\(2017\)$ begins to break down when $r\sim\rc$.
Here the formula \eqref{eq:4Aa} remains valid in the outer region with $r\gtrsim\rc$. 
So we have extended the formula for the total gravitational force to the far-away region. 
Now it is reasonable to say that the repulsive dark force will dominate over the Newtonian attraction in the outer region.

One major goal of the GODDESS mission is to detect any possible deviation from the inverse square law behavior (mentioned above) \cite{Nan18}.
It can be expected that the dark force, also referred to as `{\it the fifth force}' by the GODDESS team \cite{Nan18}, will lead to the deviation. 
It means that the GODDESS mission will have a chance of successful detection of dark energy \cite{Nan18}.
The measured force (or field) strength by GODDESS depends on its distance to the mass source.
The relative changes in the force (or field) strength has huge impact to GODDESS on differentiating the dark force from the purely Newtonian force \cite{Nan18}.
So the exact form of the total gravitational force \eqref{eq:4Aa} or the gravitational field strength \eqref{eq:gforce} can help the GODDESS team to optimize schemes 
and develop strategies for the detection of dark energy.
For example, the critical radius $\rc$ \eqref{eq:Arc} derived from equation\,\eqref{eq:4Aa} or \eqref{eq:gforce} can tell us how far the explorer should be from the Sun, the Earth or other planets, which is important to the choice of mission trajectories.

\vspace*{3mm}
\section{\hspace*{+0.0mm}Geometrization of light bending}
\label{sec:del}
\label{sec:4}
\vspace*{1mm}

In any static spacetime, there always exists a local space $\Xi_{p}$ perpendicular to the four-velocity $U^{a}$ of the static observer at each point $p$. 
The local space $\Xi_{p}$ is actually determined by a local metric tensor, denoted as $h_{ab}|_{\Xi_{p}}$, which is independent of the choice of local coordinates,
and, as seen from the local observer at point $p$, it represents a physical space. 
We can paste these local spaces $\{\Xi_{p}\}$ at different points $\{p\}$ together to get a three-dimensional manifold $\Xi$, namely $\Xi=\substack{\bigcup\\p}\,\Xi_{p}$. 
In current theories of gravity, the entire spacetime is uniquely determined by one metric tensor $g_{ab}$.
Consequently, all the metric tensors $\{h_{ab}|_{\Xi_{p}}\}$ at different points $\{p\}$ can be written in a unified form. Denote it by $h_{ab}$.
Clearly, $h_{ab}$ is the induced metric tensor from $g_{ab}$, namely $h_{ab}=g_{ab}|_{\Xi}$ (see section~\ref{sec:2} for details). 
For the static observer, as the global extension of $h_{ab}|_{\Xi_{p}}$, $h_{ab}$ is uniquely given by $g_{ab}$.
In GR, the metric tensor $g_{ab}$ can be parametrized by a set of global coordinates, like $\{z^{\mu}\}$, as shown by equation\,\eqref{eq:gab}. 
Therefore, we can describe the metric tensor $h_{ab}$ on the hypersurface $\Xi$ in the same global coordinates, exactly as demonstrated by equation\,\eqref{eq:hab} (see section~\ref{sec:2} for details).

At each point $p$, consider a locally embedded surface $\Sigma_{p}\subset\Xi_{p}$, representing a slice of $\Xi_{p}$. 
In analogy to what we have done for $\Xi$, we can paste these local surfaces $\{\Sigma_{p}\}$ at different points $\{p\}$ together
and obtain a two-dimensional manifold $\Sigma=\substack{\bigcup\\p}\,\Sigma_{p}$.
In differential geometry, $\Xi$ is a (three-dimensional) global space. 
As see from the observer at each point $p$, $\Sigma$ is a global surface embedded into the global space $\Xi$, 
with each $\Sigma_{p}$ determined by the newly induced tensor $\hat{h}_{ab}|_{\Sigma_{p}}\equiv\,\!(h_{ab}|_{\Xi_{p}})|_{\Sigma_{p}}$~$=h_{ab}|_{\Sigma_{p}}$.
Indeed, there always exists such a global surface in any curved static spacetime. 
For instance, in the SdS$_{w}$ spacetime, we can construct such a surface by setting $\theta =\frac{\pi}{2}$, without loss of generality (see section~\ref{sec:5} for details).
For any surface $\Sigma$, we denote $\hat{h}_{ab}$ to be its metric tensor. 
Generally, we have $\hat{h}_{ab}=h_{ab}|_{\Sigma}$, which is observer dependent.
In math lingo, $(\Sigma,\,\hat{h}_{ab})$ is a two-dimensional Riemannian manifold.
Now it can be viewed as a physical surface on which measurements may take place.

For each static observer, the metric tensor $\hat{h}_{ab}$, as the global extension of $\hat{h}_{ab}|_{\Sigma_{p}}$,
is not uniquely determined by $h_{ab}$. 
Actually, the tensor $\hat{h}_{ab}|_{\Sigma_{p}}$ is dependent on the choice of the local slice $\Sigma_{p}~(\subset\Xi_{p})$. 
Besides, we can always get the static observer at each point $p$ to choose a desired slice $\Sigma_{p}$ by certain rules, and then paste these local slices $\{\Sigma_{p}\}$ together to construct a global surface $\Sigma$.
To carry it out, we can define a global surface $\Sigma$ in advance. 
Then, we could have the local observer at each point $p$ choose a neighborhood of $p$ on the surface $\Sigma$ and define it as the local slice $\Sigma_{p}$. 
Clearly, the choice of $\Sigma_{p}$ is flexible, depending on how to define or choose the global surface $\Sigma$. 
In mathematics, $\hat{h}_{ab}|_{\Sigma_{p}}$ is closely related with $\Sigma_{p}$.
Thus, it also shows a dependence on the choice or the definition of $\Sigma$.
As the global extension of $\hat{h}_{ab}|_{\Sigma_{p}}$, the metric tensor $\hat{h}_{ab}$ can not be uniquely determined by $h_{ab}$; more exactly, it also depends on the choice of the global surface $\Sigma$. 
Taking into account all these factors, we come to the conclusion that the physical surface $(\Sigma,\,\hat{h}_{ab})$ can be chosen at our convenience, 
which lays a solid foundation for the geometrization of light bending.

\begin{figure}[t]
\centerline{
\includegraphics[width=1.03\columnwidth]{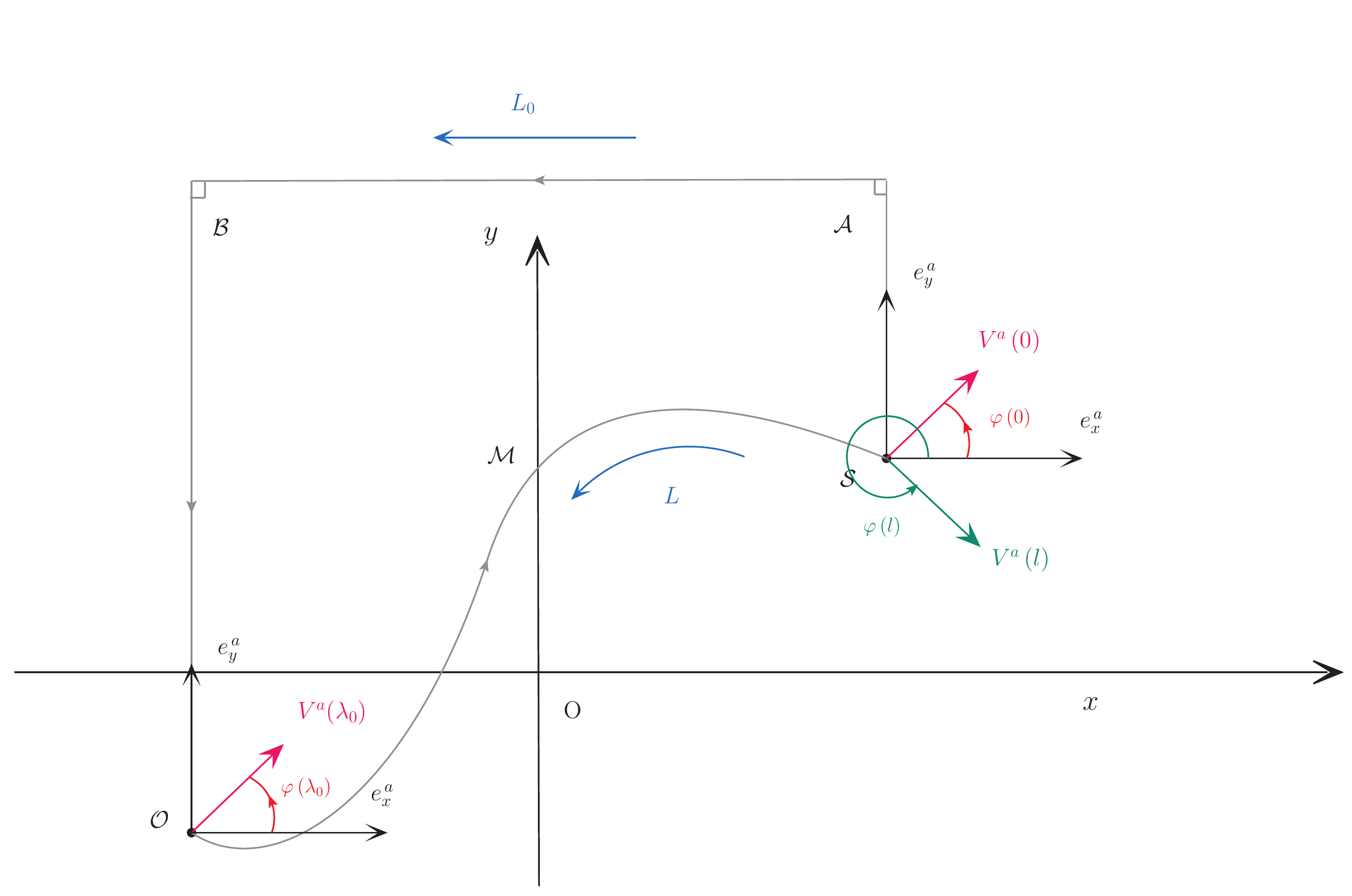}}
\vspace*{-2mm}
\caption{Parallel transport. The vector field $V^{a}\!=\!V^{a}\(\lambda\)$ denotes the parallel transport of $V^{a}\(0\)$ along the positively oriented path $\cal{S\!\to\!_{A}\!\to\!_{B}\!\to\!O\!\to_{M}\to\!S}$ at point $\lambda$, with a polar angle $\varphi\(\lambda\)$ from the $x$-axis.}
\label{fig:PTrans}
\end{figure}

Now we choose a global surface $\Sigma$. As figure~\ref{fig:PTrans} illustrates, there is a simple, closed, positively oriented piecewise regular curve on $\Sigma$,
and it is composed of two oriented paths $L_{0}\doteq\cal{S\!\to\!_{A}\!\to\!_{B}\!\to\!O}$ and $\bar{L}\doteq\cal{O\!\to_{M}\to\!S}$. 
Let $\gamma: \!I\!=\![0,l]\to\Sigma$ be a map of a closed interval $I$ to the closed curve,  with $\gamma\(\lambda_{0}\)\!=\!\cal{O}$ and $\gamma\(0\)\!=\!\gamma\(l\)\!=\!\cal{S}$, where $\lambda_{0}$ is the arc length at point $\cal{O}$, and $l$ is the total arc length of the closed curve. Here $\gamma$ has already been parametrized by arc length $\lambda$.
Now choose a Cartesian coordinate system $\rm{O}\it{xy}$ with origin at $\rm{O}=\(0, 0\)$. 
For any vector $V^{a}\(0\)$ tangent to $\Sigma$, 
let $V^{a}=V^{a}\(\lambda\)$ be the parallel transport of $V^{a}\(0\)$ along the oriented curve $\gamma$, with $\varphi\!=\!\varphi\(\lambda\)$ being the angle from the $x$-axis to $V^{a}$ at point $\lambda$. Thus, we obtain $\varphi\!=\!\varphi\(\lambda_{0}\)$ at $\lambda=\lambda_{0}$ and $\varphi\!=\!\varphi\(l\)$ at $\lambda=l$.
Along the same track as $\bar{L}$, the oriented path $L\doteq\cal{S\!\to_{M}\to\!O}$ is traced in the opposite direction. Denote $\bar{V}^{a}=\bar{V}^{a}(\bar{\lambda})$ as the parallel transport of $V^{a}\(0\)$ along $L$ at $\bar{\lambda}=l\!-\!\lambda$ and $\bar{\varphi}\!=\!\bar{\varphi}(\bar{\lambda})$ as the corresponding intersection angle between $\bar{V}^{a}$ and the $x$-axis. In this notation, we have $\bar{\varphi}\!=\!\bar{\varphi}\(\bar{\lambda}_{0}\)$ at $\bar{\lambda}_{0}=l\!-\!\lambda_{0}$. Similarly, the parallelly transported $V^{a}\(l\)$ along $L$ and the corresponding intersection angle are denoted by $\bar{V}_{*}^{a}=\bar{V}_{*}^{a}(\bar{\lambda})$ and $\bar{\varphi}_{*}\!=\!\bar{\varphi}_{*}(\bar{\lambda})$ at point $\bar{\lambda}$, respectively. In particular, $\bar{\varphi}_{*}\!=\!\bar{\varphi}_{*}\(\bar{\lambda}_{0}\)$ at $\bar{\lambda}_{0}=l\!-\!\lambda_{0}$. 
According to the theorem of uniqueness of parallel transport, we have $\bar{V}_{*}^{a}(\bar{\lambda_{0}})=V^{a}\(\lambda_{0}\)$ and $\bar{\varphi}_{*}\(\bar{\lambda}_{0}\)=\varphi\(\lambda_{0}\)$,
which are an immediate consequence of the theorem of existence and uniqueness of differential equations  \cite{Carmo16,Chern00}. 
When any two vectors are parallelly transported together along the same track and in the same direction, their intersection angle keeps to be unchanged \cite{Rindler06}.
Thus, $\bar{\varphi}_{*}\(\bar{\lambda}_{0}\)-\bar{\varphi}\(\bar{\lambda}_{0}\)=\varphi\(l\)-\varphi\(0\)$. 
Then we have $\varphi\(\lambda_{0}\)-\bar{\varphi}\(\bar{\lambda}_{0}\)=\varphi\(l\)-\varphi\(0\)$.  
It is noteworthy that these results are all independent of the choice of $V^{a}\(0\)$.
Following the same steps as above, we can derive $\varphi\(\lambda\)-\bar{\varphi}\(\bar{\lambda}\)=\varphi\(l\)-\varphi\(0\)$ for any point $\lambda$. 
Hence, we finally find the following relationship, 
\beqa
\label{eq:varphi0}
\ba{rcl}
\dis
\bar{\varphi}\(\bar{\lambda}\)-\varphi\(\lambda\)=\bar{\varphi}\(\bar{\lambda}_{0}\)-\varphi\(\lambda_{0}\),
\hspace*{0mm}
\ea
\eeqa
which holds well for any $\gamma$-like curve. 
Basing on this newly discovered relationship, we can extend the concept of light deflection to any curved static spacetime.

To do this, we need to perform the geometrization of light bending in advance.
First of all, we define a generalized deflection angle exactly as  
\beqa
\label{eq:GaussianCurvature0}
\ba{rcl}
\dis
\alpha_{M}=\bar{\varphi}\(\bar{\lambda}_{0}\)-\varphi\(\lambda_{0}\),
\hspace*{0mm}
\ea
\eeqa
which is applicable to any curved static spacetime. Hereafter, we refer to this angle as the \textit{Gaussian deflection angle}. 
Then, denote by $K$ the Gaussian curvature. By the Gauss-Bonnet theorem, we have (see appendix \ref{app:E} for details)
\beqa
\label{eq:GBtoPT}
\ba{rcl}
\dis
\varphi\(l\)-\varphi\(0\)=\int\!\!\!\int_{D}\,K\d \sigma 
\hspace*{0mm}
\ea
\eeqa
where $\d \sigma$ is the element of area and $\it{D} \(\subseteq\rm{\Sigma}\)$ denotes the simple, connected region bounded by the closed curve $\gamma$. 
Combining with the above derivations, the Gaussian deflection angle can be further derived as follows,
\beqa
\label{eq:GaussianCurvature}
\ba{rcl}
\dis
\alpha_{M}=\bar{\varphi}\(\bar{\lambda}\)-\varphi\(\lambda\)=-\int\!\!\!\int_{D}\,K\d \sigma,
\hspace*{0mm}
\ea
\eeqa
which is independent of specific spacetime models. 
It clearly indicates that the nature of light bending is the curvature of spacetime. 
Besides, the left hand side of the equation is invariant under coordinate transformations, and so is the right hand side.
Now we have geometrized the deflection of light successfully, and extended the definition of the deflection angle to the most general static spacetime.

The mathematics involved in equation\,\eqref{eq:GaussianCurvature} is fairly simple. Denote $\partial{\rm D}$ as the closed boundary of the region $D$, namely $\gamma=\partial{\rm D}$.
Choose ${\it D}$ to be a geodesic polygon (that is, polygon whose sides are arcs of geodesics). Hereafter, ${\it D}$ is referred to as the \textit{lensing patch}, on which the measurement can be made by local observers. 
In the case of gravitational lensing, light follows null geodesics.
Thus, the geodesic curvature of $\partial{\rm D}$ is zero, namely, $k_{g}\equiv0$ \cite{Carmo16,Chern00}. 
Let $\theta_{i}$ be the $i$th external angle of $\partial{\rm D}$, which is actually a measurable intersection angle (see appendix \ref{app:A} for details) by the static observer at the $i$th vertex $\gamma\(\lambda_{i}\)$.
Here $\lambda_{i}$ is the arc length at the $i$th vertex.
Then, according to the Gauss-Bonnet theorem, we also have (see appendix \ref{app:E} for details) 
\beqa
\label{eq:polygon}
\ba{rcl}
\dis
\alpha_{M}=\sum_{i}\theta_{i}-2\pi.
\hspace*{0mm}
\ea
\eeqa
Generally speaking, it asserts that the Gaussian deflection angle is equal to the excess over $2{\rm \pi}$ of the sum of the external angles of the geodesic polygon.
Take a geodesic triangle for example. Denote the $i$th interior angle as $\psi_{i}$. When dealing with the geodesic triangle, we have $\alpha_{M}=\pi-\sum_{i=1}^{3}\psi_{i}$. If the triangle is on a flat surface, we have $\alpha_{M}=0$. Also, on a sphere-like surface, $\alpha_{M}<0$ \cite{Carmo16}. Similarly, on a pseudosphere-like surface, $\alpha_{M}>0$ \cite{Carmo16}. 
Obviously, the formula \eqref{eq:polygon} provides a remarkable relation between the geodesic polygon and the deflection of light .

The Gaussian deflection angle $\alpha_{M}$ \eqref{eq:GaussianCurvature0} is actually a generalized deflection angle. In some special cases, it can reduce to the \textit{usual deflection angle}. 
To show this, we need to choose a spacetime region where the metric is flat (or conformally flat).
A typical example of this is the region of spatial infinity in the Schwarzschild spacetime. 
Hereafter, we name this kind of region as the \textit{laboratory area}.  Its properties are presented detailedly in appendix~\ref{app:A};
in the laboratory area, the light ray travels along a physically straight line, and we can not probe its bending effect.
As illustrated in figure~\ref{fig:PTrans}, let $L_{0}$ be a simple, oriented piecewise regular curve in the laboratory area.
It contains three segments of straight lines: $\,\overline{\cal{SA}}\,$, $\,\overline{\cal{AB}}$, and $\,\overline{\cal{BO}}$,
satisfying $\,\overline{\cal{SA}}\,\bot\,\(\partial/\partial\it{x}\)^{a}$, $\,\overline{\cal{AB}}\,\bot\,\(\partial/\partial\it{y}\)^{a}$, and $\,\overline{\cal{BO}}\,\bot\,\(\partial/\partial\it{x}\)^{a}$. 
These segments can be parts of light trajectories.
If so, the intersection angle between any two segments is actually an measurable angle (see appendix \ref{app:E} for details). 
Due to the flatness or the conformal flatness of the laboratory area, one has $\varphi\(\lambda_{0}\)=\varphi\(0\)$ when taking the parallel transport of $V^{a}\(0\)$ along the path $L_{0}$. 
Then, $\alpha_{M}=\bar{\varphi}\(\bar{\lambda}_{0}\)-\varphi\(0\)$, which agrees with the definition of the \textit{usual deflection angle}. 
Thus, the Gaussian deflection angle $\alpha_{M}$ recovers the usual deflection angle. 
We therefore via equations\,\eqref{eq:GaussianCurvature0} and \eqref{eq:GaussianCurvature} give a reasonable and natural generalization of the usual deflection angle.

Turn now to the application of the Gaussian deflection angle to measuring the local spacetime effect on the bending of light. To obtain the Gaussian deflection angle, 
one needs to measure the external angle $\theta_{i}$ or the interior angle $\psi_{i}$ at each vertex $\gamma\(\lambda_{i}\)$ of $\partial{\rm D}$. 
These external (or interior) angles are actually the intersection angles between any two light trajectories (as null geodesics),
and each angle can be directly measured by the static observer at each vertex. 
In reality, measurements may be performed by moving observers passing by each vertex.  
In this case, the measured values by the moving observer for $\,\theta_{i}\,$ or $\,\psi_{i}\,$ need to be made relativistic corrections.
In fact, the finally measured $\,\theta_{i}\,$ or $\,\psi_{i}\,$ by the static observer at each vertex $\gamma\(\lambda_{i}\)$ can be obtained by using the general relativistic aberration relationships \cite{Ishak-Rev}. 
Anyway, we can always obtain the Gaussian deflection angle from these locally measurable intersection angles $\,\{\theta_{i}\}\,$ or $\,\{\psi_{i}\}\,$ via the formula\,\eqref{eq:polygon}. 
In addition, the Gaussian deflection angle, as a global quantity, is fully determined by the integral of the Gaussian curvature $K$ over the lensing patch $D$, namely the total curvature (see appendix \ref{app:E} for details).
Thus, using the Gauss-Bonnet theorem, we establish a relation between the global properties of the lensing patch $D$, like the total curvature, 
and the local properties of the curve $\partial{\rm D}$ composed of null geodesics, such as $\,\theta_{i}\,$ or $\,\psi_{i}\,$ at each vertex.
By definition, $K$ quantifies properties intrinsic to the spacetime surface $\Sigma$. 
Therefore, from this perspective, the Gaussian deflection angle can be used as a potentially interesting probe of the intrinsic properties of spacetime.

\vspace*{3mm}
\section{\hspace*{+0.0mm}Local SdS$_{w}$ spacetime effect on the bending of light} 
\label{sec:5}
\vspace*{1mm}

Come back to the case of the SdS$_{w}$ spacetime. When $M=0$, the SdS$_{w}$ metric is conformally flat \cite{HZ2017}. 
In this case, the light ray travels along a physically straight line, which is the same as the Minkowski vacuum case (cf~appendix~\ref{app:A}). 
It means that in this case, we can not probe the local spacetime effect of dark energy on the bending of light. 
However, for the case of $M\neq0$, it becomes rather difficult to deal with the SdS$_{w}$ metric in a similar way. 
In fact, it is almost impossible to demonstrate what role dark energy plays in the bending of light based on the traditional theories\footnote{In traditional theories for light deflection, the usual deflection angle plays a central role, and various approaches have been proposed accordingly to probe the bending of light. Hereafter, this kind of approaches are referred to as the traditional approaches.} only
(see~appendices~\ref{app:C} and \ref{app:D} for details). 
In this section, we attempt to understand the role of dark energy in light bending by introducing the concepts and techniques presented in the previous section.
As already mentioned earlier, in the SdS$_{w}$ spacetime, there always exists a subspace $\Xi$ locally perpendicular to the four-velocity $U^{a}$ of the static observer at each point.
The subspace $\Xi$ can be globally described by the induced metric tensor, 
\beqa
\label{eq:Xi_hab}
\nonumber
h_{ab}&=&\frac{1}{1\!-2\frac{\,M\,}{r}-\!2\left(\! \frac{\,\ro\,}{r} \!\right)^{\!3w+1}}\(\d r\)_{a}\(\d r\)_{b}+ r^{2}\(\d \theta\)_{a}\(\d \theta\)_{b}+r^{2}\sin^{2}\!\theta\,\(\d\phi\)_{a}\(\d\phi\)_{b}.
\eeqa
We then take $\,\theta =\frac{\pi}{2}$, without lossing generality, and therefore obtain the following metric tensor, 
\beqa
\label{eq:Sigma_hab}
\nonumber
\hat{h}_{ab}=\dfrac{1}{1-2\frac{\,M\,}{r}-2\left(\! \frac{\,\ro\,}{r} \!\right)^{\!3w+1}}\(\d r\)_{a}\(\d r\)_{b}+ r^{2}\(\d\phi\)_{a}\(\d\phi\)_{b}. 
\eeqa
Thus the subspace $\Xi$ reduces to the $\(\it{r},\phi\)$-plane determined by $\hat{h}_{ab}$. 
We therefore construct a physical surface $\Sigma$, which is characterized by the following metric, 
\beqa
\label{eq:Sigma_SdS}
\nonumber
\d s^{2}&=&E\,\d\,\!u^{2}+2 F\,\d\,\!u\,\d\upsilon+G\,\d\upsilon^{2}=\frac{1}{1\!-2\frac{\,M\,}{r}-2\left(\! \frac{\,\ro\,}{r} \!\right)^{\!3w+1}}\, \,\d r^2\,+r^{2}\,\d\phi^2.
\eeqa
This metric is orthogonally parametrized by $\(u, \upsilon\)\!=\!\(r, \phi\)$ (see appendix \ref{app:E} for details). In this case, 
\beqa
\label{eq:EFG}
E=\frac{1}{1\!-2\frac{\,M\,}{r}-\!2\left(\! \frac{\,\ro\,}{r} \!\right)^{\!3w+1}},~F=0,~G=r^2.  
\eeqa
With these, we further derive the Gaussian curvature by using equation\,\eqref{eq:GaussK}, exactly as follows, 
\beqa
\label{eq:K_SdSw}
K=\frac{\partial}{\partial\,\!r}\(\sqrt{1\!-2\frac{\,M\,}{r}-\!2\left(\! \frac{\,\ro\,}{r} \!\right)^{\!3w+1}}\).
\eeqa
As shown in appendix \ref{app:E}, the total curvature $K_{\rm{tot}}$ can be written as
\beqa
\label{eq:totalK}
\nonumber
K_{\rm{tot}}&=&\int\!\!\!\int_{D}\,K\d \sigma
=-\frac{1}{2}\oint_{\partial\rm{D}}\(-\frac{(\sqrt{E})_{\!\upsilon}}{\sqrt{G}}\,\rm{d}\it{u}+\frac{(\sqrt{G})_{\!u}}{\sqrt{E}}\,\rm{d}\it{\upsilon}\)
\hspace*{0mm}
\eeqa
Combining this with equation\,\eqref{eq:GaussianCurvature} and equation\,\eqref{eq:EFG}, the Gaussian deflection angle can be finally reexpressed as  
\beqa
\label{eq:ralphaw}
\alpha_{M}
&=&\frac{1}{2}\,\mathlarger{\oint}_{\partial{\rm D}}\sqrt{1\!-2\frac{\,M\,}{r}-\!2\(\! \frac{\,\ro\,}{r} \!\)^{\!3w+1}}\d \phi, 
\vspace*{6mm} 
\eeqa
where $\phi=\phi\(r\)$ is uniquely determined by the boundary curve $\partial{\rm D}$. Note that the formula \eqref{eq:ralphaw} is derived without any of further assumptions or approximations.
Obviously, the Gaussian deflection angle $\alpha_{M}$ takes a model-independent form, with various dark energy models described by different $w$ values \cite{DEmodel}. 
From the formula\,\eqref{eq:ralphaw}, we can further conclude that dark energy does contribute the bending of light via the $w$-dependent term.
So it is possible to extract the information about $\(w, \ro\)$ by making precise measurements via equation\,\eqref{eq:ralphaw} for the bending of light. 
Clearly, this will have many applications.
For instance, as expected from the model predictions of the cosmological constant as the dark energy candidate, we have $w=-1$ and $\CC=6/{\ro}^{\!\!2}$,
which can be tested directly through measuring the bending of light on astrophysical scales, independently of current cosmological observations.

\begin{figure}[!htb]
  \centerline{
      \includegraphics[width=0.71\columnwidth]{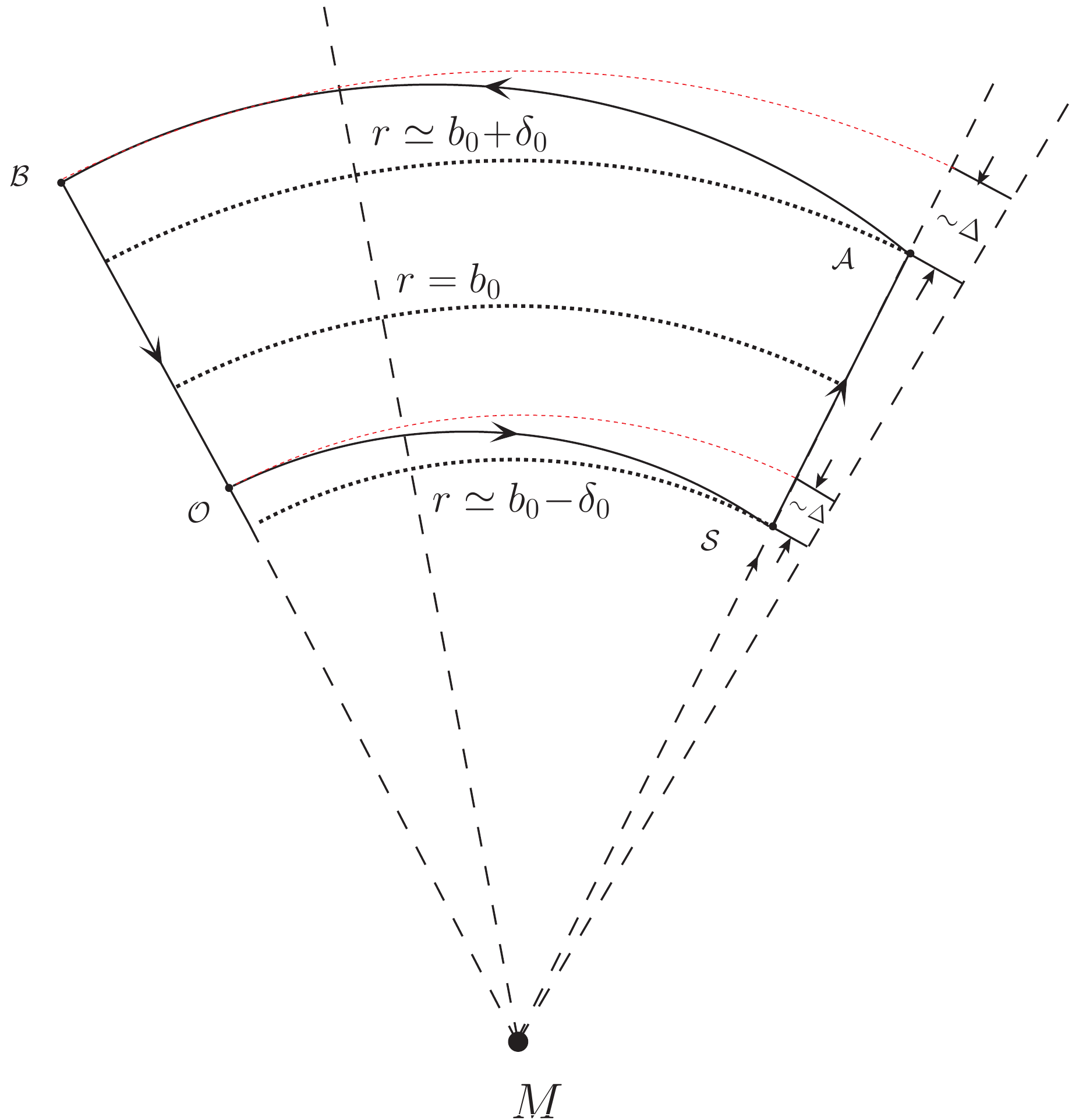}
      }
  \caption[PTV]{An example of the lensing patch. It is enclosed by a specific boundary curve, which comprises four segments of null geodesics: $\overline{\cal{SA}}$,~\,$\widetilde{\cal{AB}}$,~\,$\overline{\cal{BO}}$,~\,and~\,$\widetilde{\cal{OS}}$, with impact parameters $b=$ 0, $b_{0}+\delta_{0}$, 0, and $b_{0}-\delta_{0}$, respectively.} 
  \label{fig:BC}
\end{figure}

The boundary curve $\partial{\rm D}$ involved in the formula\,\eqref{eq:ralphaw} can be chosen at our convenience,
which is one of the significant advantages of the Gaussian deflection angle over the usual deflection angle\footnote{
This arises from the fact that the Gaussian deflection angle \eqref{eq:GaussianCurvature0} can be rewritten as a surface integral \eqref{eq:GaussianCurvature} and thus it actually has one more degree of freedom in the definition than the usual deflection angle.}.
For certain choices, we can estimate the contribution of dark energy to the Gaussian deflection angle in an analytic way.   
Now, consider a closed, positively oriented curve $\partial{\rm D}$ such that 
it consists of four segments of light trajectories: $\overline{\cal{SA}}$,~\,$\widetilde{\cal{AB}}$,~\,$\overline{\cal{BO}}$,~\,and~\,$\widetilde{\cal{OS}}$, as illustrated in figure \ref{fig:BC}. 
Their impact parameters are: $b=$ 0, $b_{0}+\delta_{0}$, 0 and $b_{0}-\delta_{0}$, respectively. 
According to the light orbit equation (LOE)\,\eqref{eq:SdSwlightLC} (see He\,\,\&\,\,Zhang~$\(2017\)$ \cite{HZ2017} for details), these segments can be described by  
\beqa
\label{eq:4lightLC}
\nonumber
\frac{\d \phi}{\d r}&=&0~~~(\overline{\cal{SA}}),\\[3mm]
\nonumber
\(\!\frac{1}{r^2}\frac{\d r}{\d \phi}\!\)^{\!\!2}&=&\frac{1}{\(b_{0}\!+\!\delta_{0}\)^2}-\frac{1}{\,r^2\,}\!\left[1\!-2\frac{M}{r}-2\!\left(\! \frac{\,\ro\,}{r}\!\right)^{\!3w+1}\right]~(\widetilde{\cal{AB}}),\\[3mm]
\nonumber
\frac{\d \phi}{\d r}&=&0~~~(\overline{\cal{BO}}), \\[3mm]
\nonumber
\(\!\frac{1}{r^2}\frac{\d r}{\d \phi}\!\)^{\!\!2}&=&\frac{1}{\(b_{0}\!-\!\delta_{0}\)^2}-\frac{1}{\,r^2\,}\!\left[1\!-2\frac{M}{r}-2\!\left(\! \frac{\,\ro\,}{r}\!\right)^{\!3w+1}\right]~(\widetilde{\cal{OS}}),
\nonumber
\vspace*{1mm} 
\eeqa
respectively. Among these segments, the ones with $b=0$ are actually parts of radial null geodesics, respectively. 
To calculate the Gaussian deflection angle, we need to perform a transformation, $\d\phi\to(\frac{\d\,\!r}{\d\phi})^{-1}\,\d\,\!r$.
With this transformation, by substituting the four LOEs into the formula\,\eqref{eq:ralphaw} and integrating it along the oriented curve $\partial{\rm D}$ as well as making some necessary approximations,
one can further provide an order-of-magnitude estimate for the Gaussian deflection angle.

To do this, we first set $\,R_{\cal{A}}<R_{\cal{B}}\,$ and $\,R_{\cal{S}}<R_{\cal{O}}\,$, exactly as portrayed in figure~\ref{fig:BC},
where $R_{\cal{P}}$ denotes the radius at which the point $\cal{P}$ is located. 
Then we use $\Delta$ to characterize half the absolute size of the change in the distance to the mass center $M$ from the photon traveling along a given path, 
such as \,$\widetilde{\cal{AB}}$\, or \,$\widetilde{\cal{OS}}$. 
By setting $\delta_{0} \ll b_{0}$, and $\Delta \ll b_{0}$, we have $r\simeq\,\!b_{0}$ for the closed curve $\partial{\rm D}$.
Thus the parameter $b_{0}$ can be used to characterize the distance of the lensing patch $D$ to the mass center $M$. 
Now we consider the gravitational potential to be fairly weak, $\frac{M}{r}\ll1$ and $\left(\! \frac{\,\ro\,}{r}\!\right)^{\!3w+1}\ll1$.
More precisely, we further set $\Delta\simeq\frac{\,R_{\cal{B}}-R_{\cal{A}}\,}{2}\sqrt{1\!-2\frac{\,M\,}{b_{0}}-\!2\left(\! \frac{\,\ro\,}{b_{0}} \!\right)^{\!3w+1}}\approx\frac{\,R_{\cal{B}}-R_{\cal{A}}\,}{2}$ for the path $\,\widetilde{\cal{AB}}\,$, 
and $\Delta\simeq\frac{\,R_{\cal{O}}-R_{\cal{S}}\,}{2}\sqrt{1\!-2\frac{\,M\,}{b_{0}}-\!2\left(\! \frac{\,\ro\,}{b_{0}} \!\right)^{\!3w+1}}\approx\frac{\,R_{\cal{O}}-R_{\cal{S}}\,}{2}$ for the path $\,\widetilde{\cal{OS}}\,$. 
Let the two paths, $\,\widetilde{\cal{AB}}\,$ and $\,\widetilde{\cal{OS}}\,$, have nearly the same $\Delta$.
As a result, the lensing patch $D$ may be a long, narrow belt. However, its length can be much smaller than $b_{0}$ in our scheme for the direct probe of dark energy.
Then we assume $\( \frac{\,\ro\,}{b_{0}}\)^{\!3w+1}\ll\frac{M}{b_{0}}$. 
Under these assumptions and approximations, we can estimate the Gaussian deflection angle for the case of $w\simeq-1$ as follows, 
\beqa
\label{eq:ApproxGDA}
\ba{rcl}
\dis
\alpha_{M}~=&&\frac{1}{2}\,\mathlarger{\oint}_{\partial{\rm D}}\frac{\sqrt{1\!-2\frac{\,M\,}{r}-\!2\(\! \frac{\,\ro\,}{r} \!\)^{\!3w+1}} }{ \(\frac{\d r}{\d \phi}\)} \d r\\[4mm]
=&&\left[~...~\right]_{\overline{\cal{SA}}}+\left[~...~\right]_{\widetilde{\cal{AB}}}+\left[~...~\right]_{\overline{\cal{BO}}}+\left[~...~\right]_{\widetilde{\cal{OS}}}\\[4mm]
\sim&&0+\frac{\frac{\Delta}{b_{0}\!+\!\delta_{0}}}{\sqrt{\frac{M}{b_{0}\!+\!\delta_{0}}+\(\frac{b_{0}\!+\!\delta_{0}}{\ro}\)^2}}
+0-\frac{\frac{\Delta}{b_{0}\!-\!\delta_{0}}}{\sqrt{\frac{M}{b_{0}\!-\!\delta_{0}}+\(\frac{b_{0}\!-\!\delta_{0}}{\ro}\)^2}}\\[6mm]
\sim&&0+\frac{\Delta}{b_{0}}\sqrt{\frac{b_{0}}{M}}\left[1-\frac{1}{2}\frac{b_{0}}{M}\(\frac{b_{0}}{\ro}\)^2\right]\(1-\frac{1}{2}\frac{\delta_{0}}{b_{0}}\)\\[4mm]
+&&0-\frac{\Delta}{b_{0}}\sqrt{\frac{b_{0}}{M}}\left[1-\frac{1}{2}\frac{b_{0}}{M}\(\frac{b_{0}}{\ro}\)^2\right]\(1+\frac{1}{2}\frac{\delta_{0}}{b_{0}}\)\\[4mm]
=&&\(\frac{\Delta}{b_{0}}\)\(\frac{\delta_{0}}{b_{0}}\)\sqrt{\frac{b_{0}}{M}}\left[1-\frac{1}{2}\frac{b_{0}}{M}\(\frac{b_{0}}{\ro}\)^{2}\right],
\ea
\eeqa
where $\left[~...~\right]_{\rm{part}}$ represents the line integral~\eqref{eq:ralphaw} along a part of the boundary curve $\partial{\rm D}$. 
Let's think about this line by line. The first line comes directly from equation\,\eqref{eq:ralphaw}. 
In the second line, the four LOEs have been substituted into $\frac{\d\,\!r}{\,\,\d\phi\,\,}$, respectively. 
As expected, this line allows simply setting $M=0$ and finding that the Gaussian deflection angle is zero to a first approximation.
The third line uses the two approximate formulae of $\Delta$ (mentioned above) for the two paths, $\,\widetilde{\cal{AB}}\,$ and $\,\widetilde{\cal{OS}}$, respectively. 
The fourth line has been derived by using $\delta_{0} \ll b_{0}$ and $(\! \frac{\,\ro\,}{b_{0}}\!)^{\!3w+1}\ll\frac{M}{b_{0}}$.  
The last line gives an approximate formula quite roughly, but this will suffice for an order-of-magnitude estimate.
In the last line, the Gaussian deflection angle takes a simple form, providing a straightforward way to understand the role of dark energy in the bending of light.

For comparison, we also calculate the \textit{usual deflection angle} by using some techniques developed for the Gaussian deflection angle (see appendix \ref{app:D} for details). 
For the special case of the cosmological constant ($w=-1$), the \textit{usual deflection angle} can be approximated as
\beqa
\label{eq:UDA4w=-1}
\alpha_{M}\sim\,\frac{4M}{b} \left[1+\(\frac{b}{r_{o}}\)^2\right]=\frac{4M}{b}+\frac{2M\CC}{3}\,b,
\eeqa
where the $\CC$ term is consistent with the result presented in the literature \cite{Italy,Sereno08,Bhadra10,Ishak-Rev,Arakida12}. 
When setting $\CC=0$, we find that the usual deflection angle recovers the conventional result of light bending, $\alpha_{M}=\frac{4M}{b}$. 
We can derive the dark energy correction $\Delta\alpha_{M}$ to the usual deflection angle,  
\beqa
\label{eq:approxSDA}
\frac{\Delta\alpha_{M}}{\alpha_{M}}\sim\(\frac{b}{r_{o}}\)^{2}.
\eeqa
Accordingly, we find that effects of $\CC$ on the bending of light are quite small for a real astrophysical system.
For example, let $b$ equal the size of our Solar System, like $b=$ 0.1 $\rc$. Then we have $\frac{\Delta\alpha_{M}}{\alpha_{M}}\sim10^{-18}$ from equation\,\eqref{eq:approxSDA}. 
Similar to Sereno (2019)\,\cite{Italy,Sereno08}, we can conclude from the point of view of the traditional (gravitational lensing) theories that effects due to $\CC$ are too small to be detected, 
although the $\CC$ term does contributes to the usual deflection angle.

From the formula \eqref{eq:ApproxGDA}, 
we also get the dark energy correction to the Gaussian deflection angle, 
\beqa
\label{eq:relativeDA}
\frac{\Delta\alpha_{M}}{\alpha_{M}}\sim\frac{1}{2}\frac{\,\,b_{0}\,\,}{\,\,M\,\,}\(\frac{b_{0}}{\ro}\)^{\!2}.
\eeqa
Clearly, it is about $\sim\frac{1}{2}\(\!\frac{\,\,b_{0}\,\,}{\,\,M\,\,}\!\)$ times larger than that to the usual deflection angle. 
It is worth rephrasing that we can further enhance the local effect of dark energy via the proper choice of the boundary curve $\partial{\rm D}$.

In our Solar System, we have $M\sim1.0 \,M_{\odot}$, where $M_{\odot}$ is the mass of the Sun. 
Set $b_{0}=$ 0.1 $\rc$ and $\delta_{0}\sim\Delta\sim1.0~\textrm{AU}$ (astronomical unit). In this case, it can be verified that all the assumptions and approximations hold well throughout the derivations of the formula\,\eqref{eq:ApproxGDA}.
Then, $\frac{1}{2}\(\!\frac{b_{0}}{\,\,M\,\,}\!\)\sim2.2\times10^{14}$.
It clearly indicates that the influence of dark energy on light bending can be enhanced by 14 orders of magnitude. 
Notice that in our scheme, the impact parameter is fixed to a design value, and it is not an observable.
From equations\,\eqref{eq:ApproxGDA} and \eqref{eq:relativeDA}, we further obtain $\alpha_{M}\sim0.63''$ and $\frac{\Delta\alpha_{M}}{\alpha_{M}}\sim2.5\times10^{-4}$. 
It means that we can directly probe the existence of dark energy and measure the equation-of-state parameter $w$ on a much shorter length scale than the Solar-System's 
at the distance of $\sim$ 0.1 $\rc$ from the Sun once a spatial resolution of $\sim\!10^{-5}$ arcseconds can be reached by the current lensing experiments. 
In fact, this spatial resolution has been achieved by GRAVITY \cite{GRAVITY18}, 
which is often used to detect the gravitational microlensing events.

The universe is presently dominated by not only dark energy, but also the pressure-less (baryonic and dark) matter.
However, when considering the fact that the Sun contains more than 99\% of the mass of the Solar System,
as well as the SdS$_{w}$ metric holds well for a point-like mass $M$ or for regions outside a spherically symmetric mass-distribution \cite{HZ2017}, 
it can be found that our estimate remains valid, especially as an order-of-magnitude estimate.
So we can safely draw the conclusion that the current lensing experiments are already sensitive to probing dark energy via the method presented above.

Nevertheless, these lensing experiments are all designed based on the traditional approaches where 
it is the usual deflection angle that plays a central role rather than the Gaussian deflection angle \eqref{eq:GaussianCurvature}. 
Within the framework of traditional theories, the local dark energy effect is too weak to be detected only by using traditional approaches (as we discussed just below equation\,\eqref{eq:approxSDA} or shown by Sereno (2019) \cite{Italy,Sereno08} for the $\CC$ effects, for example); 
in other words, we have no chance to amplify the local effect of dark energy via the choice of the path of integration $\partial{\rm D}$
in analogy to what we have done for the Gaussian deflection angle \eqref{eq:ralphaw}.
This explains why despite intense effort no experiment on the bending of light shows any deviation from traditional theories, till today.
Anyway, we have successfully proposed a method to overcome the difficulty of measuring local dark energy effects.

\vspace*{4mm}
\section{\hspace*{+0.0mm}Conclusions}
\label{sec:6}
\vspace*{1.5mm}

Recently, the XENON1T reported a non-gravitational signature of dark energy \cite{Vagnozzi21}.
If it is confirmed by future experiments, 
we have to face the scenario in which dark energy couples to photons directly. 
Taking dark energy as a scalar field for example, its quanta interact with photons, as assumed in \cite{Vagnozzi21}. 
If so, these quanta may be produced by the magnetized objects like the Sun and then propagate outward to the observer, forming local fluctuations in the pressure and energy density of dark energy
while still keeping $w$ nearly constant on astrophysical scales so as not to contradict with the existing observational data. 
These fluctuations may manifest as tiny noises or perturbations when measuring the local dark energy effect via the bending of light. 
On the other hand, in principle this scenario can only be dealt with in quantum gravity. 
As was done in \cite{QGbending15,Bai17}, new scalar fields were introduced into the standard model of particle physics, 
and their quanta interact with photons at the quantum level.
It turns out that the final result of quantum gravity agrees with the classical GR result to a first-order approximation \cite{Bai17}.
However, when returning back to the study of dark energy, it might become rather different to calculate the deflection angle, 
and the computations need to be accomplished strictly using modern field theory techniques, which is far beyond the scope of our work.

In this work, we focus on the classical GR, and showed that the bending of light can serve as an important tool for the direct probe of dark energy on the Solar System scales. 
By using the famous Gauss-Bonnet theorem, we geometrized light deflection, and demonstrated explicitly that in any curved static spacetime the Gaussian deflection angle is equivalent to the total curvature.  
For the general case of the SdS$_{w}$ spacetime, we concluded that dark energy does affect the deflection of light. Measuring such effect can directly probe the existence and nature of dark energy.

In section~\ref{sec:2}, we introduced the projection tensor $h^{a}_{\,\,\,b}$ and the induced metric tensor $h_{ab}$. 
Both of them are observer dependent.
Generally, $h^{a}_{\,\,\,b}$ projects any four-vector onto the observer's local space that is described by the metric tensor $h_{ab}$. 
In physics, the projection of the four-vector by $h^{a}_{\,\,\,b}$ is physically measurable. 
For example, the proper three-acceleration $\hat{a}^{a}$ measured by the static observer can be regarded as the projection of the four-acceleration $\hat{A}^{a}$. More exactly, we have $\hat{a}^{a}=- h^{a}_{\,\,\,b}\hat{A}^{b}$~ $(=-\hat{A}^{a})$, with $\pi^{a}_{\,\,\,b}\,\hat{A}^{b}\!=\!0$. It is independent of the choice of coordinates, since all the Lorentz indices are contracted. This is indeed the reason why the gravitational three-force $\vec{g}$ can be directly measured by the static observer in the SdS$_{w}$ spacetime.

In section~\ref{sec:3}, we derived the proper form for the dark force induced by dark energy, in analogy to what we have done for the traditional Newtonian gravity.
Then we further showed that the dark force is repulsive and its strength grows with $r$. This is in contrast to the attractive Newtonian force caused by both visible and dark matter.
We also derived the critical radius $\rc$, at which the dark force balances the Newtonian attraction.
$\rc$ plays an important role in helping us to design experiments and develop strategies for the direct probe of dark energy through gravitational effects.  
Taking the gravitational lensing effect for example, the lensing patch $D$ cannot be too far away from the region $r\sim\rc$, otherwise it will become very difficult for the direct probe of dark energy.
Another example is about the GODDESS mission, which has been proposed by NASA to detect the dark force (also referred to as the fifth force). 
Using the explicit form of the dark force or its strength, we can maximally optimize detection schemes and quantitatively develop measurement strategies for the GODDESS mission.

Section~\ref{sec:4} was built on the Gauss-Bonnet theorem. 
We first defined the Gaussian deflection angle by adopting new techniques, like taking parallel transport. 
Using the Gauss-Bonnet theorem, we then proved strictly that the Gaussian deflection angle is identical to the integral of the Gaussian curvature over the lensing patch $D$, 
which is applicable in any curved static spacetime.  
We therefore geometrized the bending of light after a sequence of strict mathematical derivations.
We also showed that the Gaussian deflection angle is intrinsically a generalized deflection angle;
in some special cases, the Gaussian deflection angle reduces to the usual deflection angle.
From the Gauss-Bonnet theorem, we also derived a relationship between the Gaussian deflection angle and the measured external (or interior) angles 
by the static observers at every vertex of the closed boundary curve $\partial{\rm D}$. 
From this relationship, the Gaussian deflection angle can be obtained directly.
Note that all these results are independent of the choice of coordinates, forming a strict mathematical basis for the theories of direct probe of dark energy through the bending of light.

In section~\ref{sec:5}, we demonstrated that the light ray is deflected as it travels in the SdS$_{w}$ spacetime, 
and gained a clear understanding of the role of dark energy in the bending of light. 
For the Gaussian deflection angle, we did explicit calculations in the general SdS$_{w}$ case (cf appendix \ref{app:E}), and presented its analytical form in equation\,\eqref{eq:ralphaw} strictly; 
the deflection angle takes a general form, where the equation-of-state parameter $w$ can describe different forms of dark energy when it takes their corresponding values. 
Thus we model-independently concluded that dark energy does affect the deflection of light. 
We also showed in equation\,\eqref{eq:ralphaw} that the dark energy contribution is fully determined by the two generic parameters $\(w, \ro\)$. 
Therefore, it is possible to obtain $\(w, \ro\)$ from measuring the Gaussian deflection angle.
With extracted information about the two parameters, we could further discriminate between different dark energy models and thus identify the right model for describing dark energy.

In addition, we proposed a method to overcome the difficulty of measuring the local effect of dark energy on the bending of light.
Under the weak field approximation, we calculated the direct contribution of dark energy to the Gaussian deflection angle,
trying to maximize the effect of dark energy by choosing the lensing patch $D$. 
We also estimated the usual deflection angle by using the techniques developed in section~\ref{sec:4} (See appendices \ref{app:A}, \ref{app:B}, \ref{app:C}, and \ref{app:D} for details). 
By comparisons, we concluded that the direct contribution of dark energy to light deflection can be enhanced by 14 orders of magnitude if we choose the boundary curve $\partial{\rm D}$ properly.
When applying this enhancement to detect dark energy in our Solar System, we found that the direct contribution of dark energy is already sensitive to the current lensing measurements.

Finally, we conclude that in general, it is important to make precise measurements on the bending of light via the Gaussian deflection angle \eqref{eq:GaussianCurvature},
and this will allow us to probe the existence of dark energy directly and discriminate different dark energy models with $w=-1$ versus $w\neq-1$.
By the theories established in this work, it can be expected that the direct probe of dark energy can been achieved at much shorter scales than our Solar-System's in the near future. 

\acknowledgments

The author thanks Prof. SL Xiong for his support and Prof. HJ He for useful discussions as well as the two anonymous reviewers for their valuable comments and constructive suggestions. 
This work is supported by the National Program on Key Research and Development Project (Grant No. 2016YFA0400800, 2016YFA0400802, 2017YFA0402600) from the Minister of Science and Technology of China (MOST), the Key Research Program of Frontier Sciences of the Chinese Academy of Sciences  (Grant No. QYZDY-SSW-SLH008), the Strategic Priority Research Program on Space Science of the Chinese Academy of Sciences (Grant No. XDB23040400), and the National SKA Program of China (Grant No. 2020SKA0120300). The authors thank supports from the National Natural Science Foundation of China under Grants U1838202, 11833003, U1838201, U2031205, 11673023, 11733009, U1838111, U1838113, U1838105, 11503027, and U1838104.

\appendix

\vspace*{3mm}
\section{\hspace*{+0.0mm}Laboratory area}	
\label{app:A}


The SdS$_{w}$ metric with $M=0$ is conformally flat\,\cite{HZ2017}. 
For the general case of $M\not=0$, in the region with $\,r\gg\rc$, 
the Newtonian attraction of matter can be ignored to a great extent, and the gravitational force is dominated by the local repulsion from dark energy (see section~\ref{sec:3} for details). 
When neglecting the Newtonian term completely, we can rewrite the SdS$_{w}$ metric under coordinate transformations as 
$\d S^{2}=\Omega^{2}(\bar{\it{r}}) (- \rm{d}\bar{\tau}^{2}+\rm{d}\bar{\it{r}}^{2}+\bar{\it{r}}^{2} \rm{d}\theta^{2}+\bar{\it{r}}^{2} sin^{2} \theta \,\rm{d}\phi^{2})$ (cf appendix\,\ref{app:C}). 
In this ideal case, dark energy has no measurable effect on the usual light deflection \cite{HZ2017}.
Hereafter, we name this kind of region as the \textit{laboratory area}.
Without losing generality, we confine the motion to the $\(\bar{\it{r}},\phi\)$-plane of $\,\theta =\frac{\pi}{2}$. Then the metric reduces to 
\beqa
\label{eq:pi/2dS^2}
\ba{rcl}
\dis
\d S^{2}=\Omega^{2}(\bar{\it{r}}) (- \rm{d}\bar{\tau}^{2}+\rm{d}\bar{\it{r}}^{2}+\bar{\it{r}}^{2} \,\rm{d}\phi^{2}).
\ea
\eeqa

In any given coordinates $\eta^{\mu}$, the energy-momentum four-vector of a light ray can be defined by $K^{a}=\frac{\,\,\,\d \eta^{\mu}}{\d\lambda}\(\!\frac{\partial\,}{\,\,\partial \eta^{\mu}}\!\)^{\!a}$, where $\lambda$ is an affine parameter. According to the metric~\eqref{eq:pi/2dS^2}, we obtain two Killing vectors $\xi^{a}=(\partial/\partial \bar{\tau})^{a}$ and $\zeta^{a}=(\partial/\partial \phi)^{a}$. This means that the metric \eqref{eq:pi/2dS^2} respects the symmetries of time translation and space rotation. Therefore, the energy $E\!=-\,\!\xi^{a}K_{a}$ and the angular momentum $L\!=\!\zeta^{a}K_{a}$ are conserved, respectively. So we have
\beqa
\label{eq:enanEqC5a}
\dis
\nonumber
E &=& \Omega^{2}\,\frac{\d \bar{\tau}}{\d \lambda}={\rm Constant} \,,~~~L=\bar{\it{r}}^2 \,\Omega^{2}\,\frac{\d \phi}{\,\d \lambda\,}={\rm Constant} ,
\eeqa
where $\Omega=\Omega\(\bar{\it{r}}\)$. Here, $E$ and $L$ are both physical quantities. Combining with the null condition $\rm{d}\bar{\tau}^{2}=\rm{d}\bar{\it{r}}^{2}+\bar{\it{r}}^{2} \,\rm{d}\phi^{2}$,
we obtain 
\beqa
\label{eq:loe_CF2}
\dis
\(\frac{1}{\bar{\it{r}}^2}\frac{\d \bar{\it{r}}}{\d \phi}\)^{2}=\frac{1}{b^2}\!-\!\frac{1}{\,\bar{\it{r}}^2\,},
\eeqa
with $b\!=\!\it{L/E}$.  Like $E$ and $L$, the impact parameter $b$ is also a physical quantity.
Furthermore, we rederive equation\,\eqref{eq:loe_CF2} as follows,
\beqa
\label{eq:straightLC5}
\ba{rcl}
\dis
\bar{\it{r}}=\frac{b}{\rm{sin}\,(\phi\!-\!\phi_{0})},
\ea
\eeqa
where $\phi$ denotes the polar angle measured in a counterclockwise direction from a given $\it{x}$-axis (as the polar axis). 
Besides, observe from equation\,\eqref{eq:straightLC5} that it describes a straight line, and $\phi_{0}$ is the intersection angle between this line and the $\it{x}$-axis.
It tells us that the light ray travels along a straight line in any conformally flat space.

\subsection{Intersection angle}
\label{subsec:IA}

Denote $A^{a}$ and $B^{a}$ as four-momenta of two intersecting light rays passing through the point of a static observer, respectively.
For the observer, the \textit{measurable intersection angle} $\angle\!(\!A\!, \!B\!)\!_{\textit{M}}$ between the two vectors is given by
\beqa
\label{eq:thetaM}
\dis
{\rm cos}\,\angle\!(\!\textit{A}\!, \!\textit{B}\!)\!_{\textit{M}} = \frac{A_{\perp \,a}B_{\perp}^{\,\,\,a}}{\big|\!A_{\perp}\!\big|\,\big|\!B_{\perp}\!\big|}
\eeqa
From equation\,\eqref{eq:pi/2dS^2}, we obtain the velocity $U^{a}$ of the static observer and its dual vector $U_{a}$, exactly as follows,
\beqa
\label{eq:staticV4}
\ba{rcl}
\dis
\nonumber
U^{a}=U^{t} \(\partial/\partial\,\!t\)^{a}=\frac{1}{\Omega}\(\partial/\partial\,\!t\)^{a},~~~U_{a}=U_{t} \,(\d\,\!t)_{a}=-\Omega \,(\d\,\!t)_{a}.
\ea
\eeqa
Then we have $h_{ab}\!=\!\Omega^{2}\left[\(\rm{d}\bar{\it{r}}\)_{a}\(\rm{d}\bar{\it{r}}\)_{b}\!+\!\bar{\it{r}}^{2} \,\(\rm{d}\phi\)_{a}\(\rm{d}\phi\)_{b}\right]$.  
Using equations \eqref{eq:Vperp}, \eqref{eq:gab}, \eqref{eq:Va} and \eqref{eq:Vperp1}, we get
\beqa
\label{eq:Aa4}
\ba{rcl}
\dis
\nonumber
A_{\perp}^{\,\,\,a}&=&A^{\bar{\it{r}}} (\partial/\partial\,\!\bar{\it{r}})^{a}+A^{\phi} (\partial/\partial\,\!\phi)^{a},~~~{\rm and}~~~
A_{\perp\,\!a}=\Omega^2\left[A^{\bar{\it{r}}} \(\rm{d}\bar{\it{r}}\)_{a}+\bar{\it{r}}^{2} A^{\phi} \(\rm{d}\phi\)_{a}\right].
\ea
\eeqa
From equation\,\eqref{eq:pi/2dS^2}, we can also obtain the tetred $\{{\textit{e}_{\mu}^{\,a}}\}$ as follows,
\beqa
\dis
\label{eq:emua1}
\textit{e}_{\bar{\tau}}^{\,a}&=&\frac{1}{\Omega}\,(\partial/\partial\,\!\bar{\tau})^{a},~~~\textit{e}_{\bar{\it{r}}}^{\,a}=\frac{1}{\Omega}\,(\partial/\partial\,\!\bar{\it{r}})^{a},~~~\textit{e}_{\phi}^{\,a}=\frac{1}{\bar{\it{r}}\,\Omega} (\partial/\partial\,\!\phi)^{a}\\[1mm]
\label{eq:emua1b}
\textit{e}^{\bar{\tau}}_{\,a}&=&-\,\Omega\,(\d \bar{\tau})_{a},\,~~~~\textit{e}^{\bar{\it{r}}}_{\,a}=\Omega\,(\d \bar{\it{r}})_{a},~~~~~~~~\textit{e}^{\phi}_{\,a}=\bar{\it{r}}\,\Omega\,(\d \phi)_{a}
\eeqa
In the tetred, we can rewrite $A_{\perp}^{\,\,\,a}$ and $A_{\perp\,a}$ as
\beqa
\label{eq:Aa_e4}
\ba{rcl}
\dis
\nonumber
A_{\perp}^{\,\,\,a}&=&\(\Omega A^{\bar{\it{r}}}\) \textit{e}_{\bar{\it{r}}}^{\,a}+(\bar{\it{r}}\,\Omega A^{\phi}) \textit{e}_{\phi}^{\,a}~~~{\rm and}~~~
A_{\perp\!a}=\(\Omega A_{\bar{\it{r}}}\)\,\textit{e}^{\bar{\it{r}}}_{\,a}+(\bar{\it{r}}\,\Omega A_{\phi})\,\textit{e}^{\phi}_{\,a},
\ea
\eeqa
respectively. It is similar for the case of $B_{\perp}^{\,\,\,a}$ and $B_{\perp\,a}$.
Then we have the measurable intersection angle between the two vectors $A^{a}$ and $B^{a}$,
\beqa
\label{eq:thetaM2d}
\ba{rcl}
\dis
{\rm cos}\,\angle\!(\!\it{A}\!, \!\it{B}\!)\!_{\textit{M}} =\frac{A^{\bar{\it{r}}}B^{\bar{\it{r}}}+\bar{\it{r}}^{\rm{2}}A^{\phi}B^{\phi}}{\sqrt{A^{\bar{\it{r}}}A^{\bar{\it{r}}}+\bar{\it{r}}^{\rm{2}}A^{\phi}A^{\phi}}\sqrt{B^{\bar{\it{r}}}B^{\bar{\it{r}}}+\bar{\it{r}}^{\rm{2}}B^{\phi}B^{\phi}}}.
\ea
\eeqa

By setting $\Omega\equiv\,\!1$, we obtain a tetred $\{E_{\mu}^{\,a}\}$ from equations\,\,\eqref{eq:emua1} and \,\eqref{eq:emua1b}, for the $\(\bar{\tau}, \bar{\it{r}}, \phi\)$-space, exactly as follows,  
\beqa
\dis
\label{eq:Emua2}
E_{\bar{\tau}}^{\,a}&=&(\partial/\partial\,\!\bar{\tau})^{a},~~~~E_{\bar{\it{r}}}^{\,a}=(\partial/\partial\,\!\bar{\it{r}})^{a},~~~~E_{\phi}^{\,a}=\frac{1}{\bar{\it{r}}} (\partial/\partial\,\!\phi)^{a}\\
\label{eq:Emua2b}
E^{\bar{\tau}}_{\,a}&=&-\,(\d \bar{\tau})_{a},~~~~~E^{\bar{\it{r}}}_{\,a}=\,(\d \bar{\it{r}})_{a},~~~~~~~~E^{\phi}_{\,a}=\bar{\it{r}}\,\,(\d \phi)_{a}
\eeqa
which is also the tetrad of a Minkowski spacetime. 
To further understand the coordinate angle in the metric, we need to define the \textit{Euclidean intersection angle} explicitly. Now we reexpress $A_{\perp}^{\,\,\,a}$ and $A_{\perp\,a}$ in the tetred $\{E_{\mu}^{\,a}\}$ as
\beqa
\label{eq:Aa_E4}
\ba{lcl}
\dis
\nonumber
A_{\perp}^{\,\,\,a}\!&=&\!\(A^{\bar{\it{r}}}\) E_{\bar{\it{r}}}^{\,a}+\(\bar{\it{r}}A^{\phi}\) E_{\phi}^{\,a}, ~~~{\rm and}~~~A_{\perp\!a}\!=\!\(A_{\bar{\it{r}}}\)E^{\bar{\it{r}}}_{\,a}+\(\bar{\it{r}}A_{\phi}\)E^{\phi}_{\,a}.
\ea
\eeqa
Similarly, one can get the exact form of $B_{\perp}^{\,\,\,a}$ or $B_{\perp\,a}$.
Accordingly, the Euclidean intersection angle $\angle\!(\!\textit{A}\!, \!\textit{B}\!)\!_{\textit{E}}$ can be expressed by
\beqa
\label{eq:thetaE2d}
\ba{rcl}
\dis
{\rm cos}\,\angle\!(\!\textit{A}\!, \!\textit{B}\!)\!_{\textit{E}} =\frac{A^{\bar{\it{r}}}B^{\bar{\it{r}}}+\bar{\it{r}}^{2}A^{\phi}B^{\phi}}{\sqrt{A^{\bar{\it{r}}}A^{\bar{\it{r}}}+\bar{\it{r}}^{2}A^{\phi}A^{\phi}}\sqrt{B^{\bar{\it{r}}}B^{\bar{\it{r}}}+\bar{\it{r}}^{2}B^{\phi}B^{\phi}}}.
\ea
\eeqa
Comparing equation\,\eqref{eq:thetaM2d} with equation\,\eqref{eq:thetaE2d}, we have 
\beqa
\label{eq:angleME2d}
\ba{rcl}
\dis
{\rm cos}\,\angle\!(\!\textit{A}\!, \!\textit{B}\!)\!_{\textit{E}} ={\rm cos}\,\angle\!(\!\textit{A}\!, \!\textit{B}\!)\!_{\textit{M}}. 
\ea
\eeqa
So the Euclidean intersection angle equals the measurable intersection angle in any conformally flat spacetime.

\subsection{Coordinate angle}
\label{subsec:CA}

It can be proved that in a conformally flat spacetime, the coordinate angle $\phi$ in the metric \eqref{eq:pi/2dS^2} is a Euclidean intersection angle, and thus a measurable intersection angle. To show this, define two radial vectors $A^{a}$ and $B^{a}$, with $A^{\phi}$ and $B^{\phi}$ being zero: $A_{\perp}^{\,\,\,a}=(\Omega\,A^{\bar{\it{r}}})\, \textit{e}_{\bar{\it{r}}}^{\,a}$ and $B_{\perp}^{\,\,\,a}=(\Omega\,B^{\bar{\it{r}}}) \,\textit{e}_{\bar{\it{r}}}^{\,a}$. Let $x=\bar{\it{r}} \,\,\rm{cos}\,\phi$ and $y=\bar{\it{r}}\,\,\rm{sin}\,\phi$. Then we can reexpress the tatred $\{\textit{e}_{\bar{\it{r}}}^{\,a},\textit{e}_{\phi}^{\,a}\}$ in the $\(\textit{x}, \textit{y}\)$-coordinates as
\beqa
\label{eq:xytatred}
\ba{rcl}
\dis
\nonumber
\textit{e}_{\bar{\it{r}}}^{\,a}\!=\! \rm{cos}\,\phi\,\textit{e}_{\textit{x}}^{\,\textit{a}} +\rm{sin}\,\phi\,\textit{e}_{\textit{y}}^{\,\textit{a}},~~~{\rm and}~~~\textit{e}_{\phi}^{\,\textit{a}}\!=\!-\rm{sin}\,\phi\,\,\textit{e}_{\textit{x}}^{\,\textit{a}}  +\rm{cos}\,\phi\,\textit{e}_{\textit{y}}^{\,\textit{a}}, 
\ea
\eeqa
with $\textit{e}_{x}^{\,a}\!=\!\frac{1}{\Omega}\(\partial/\partial\,\!\textit{x}\)^{a}$ and $\textit{e}_{y}^{\,a}\!=\!\frac{1}{\Omega}\(\partial/\partial\,\!\textit{y}\)^{a}$.
Let $A^{a}$ align with the $\it{x}$-axis and $B^{a}$ be the radial vector with a polar angle of $\phi$ from the $\it{x}$-axis. Similar to the derivation of equation\,\eqref{eq:angleME2d}, we have $\rm{cos}\,\angle\!(\!\textit{A}\!, \!\textit{B}\!)\!_{\textit{E}} \!=\! \rm{cos}\,(\phi)\!=\!\rm{cos}\,\angle\!(\!\textit{A}\!, \!\textit{B}\!)\!_{\textit{M}} $. Thus the coordinate angle $\phi$ in the SdS$_{w}$ metric is identical to the measurable intersection angle between the two radial vectors $A^{a}$ and $B^{a}$; that is, $\phi\!=\!\angle\!(\!\textit{A}\!, \!\textit{B}\!)\!_{\textit{M}}$. Note that the angles $\phi_{\rm{i}}$ and $\phi_{0}$ mentioned earlier are of the same kind; they are both measurable intersection angles. It means that in the conformally flat spacetime, the light ray travels along a physically straight line, and one can not probe its bending effect only by using the traditional approaches (see appendix~\ref{app:D} for details about the traditional approaches).

\vspace*{3mm}
\section{\hspace*{+0.0mm}Symmetric light orbit}
\label{app:B}
\vspace*{1mm}

\begin{figure}[!htb]
  \centerline{
      \includegraphics[width=1.01\columnwidth]{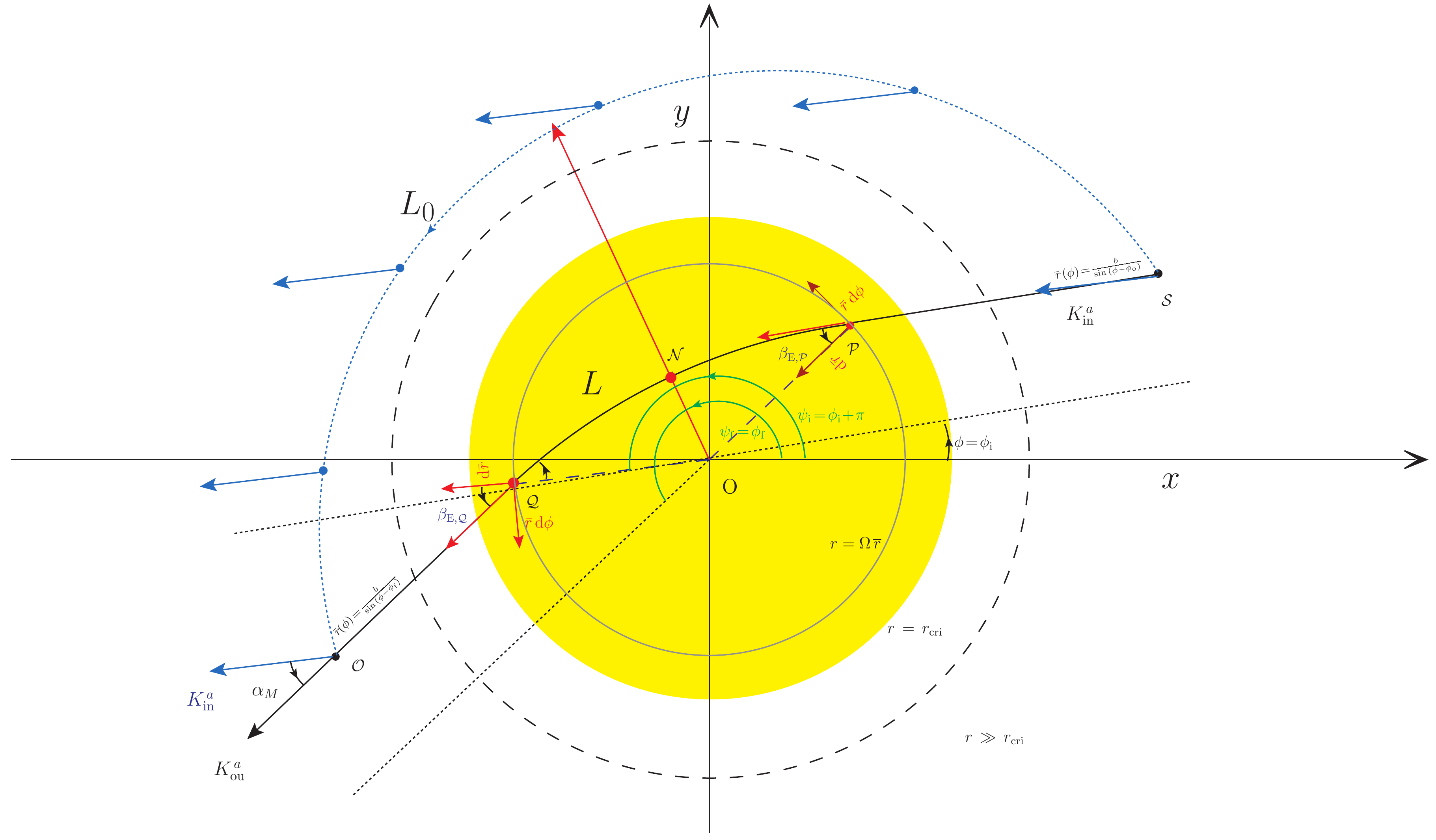}
      }
  \caption[Deflection]{Light bending. The black line $L\doteq\cal{S}\!\to\!_{\cal{N}}\!\to\!\cal{O}$ represents the path of a light ray in the $\(\it{r}, \phi\)$-plane, which is deflected by the gravitational field of matter and dark energy. The yellow area within the critical radius $r=\rc$ illustrates the region where the gravitational force is dominated by the Newtonian attraction. In this area, the circle of radius $r$ intersects with the light path $L$ at points $\cal{P}$ and $\cal{Q}$. Specifically, $\cal{N}$ is the point of closest approach of the path $L$, and $\beta_{E,\cal{P}}$\,$\big(\beta_{E,\cal{Q}}\big)$ is the Euclidean intersection angle between the radial direction and the three-momentum of the ray at point $\cal{P}$\,$\big(\cal{Q}\big)$. Outside the dashed circle, it is the outer region with $r\gg\rc$. The two points $\cal{S}$ and $\cal{O}$ are both located in this region. The vector $K_{\rm{in}}^{\,a}$ with $\phi=\phi_{\rm{i}}$ represents the four-momentum of  the incident ray at point $\cal{S}$, and $K_{\rm{ou}}^{\,a}$ with $\phi=\phi_{\rm{f}}$ that of the outgoing ray at point $\cal{O}$. 
  }
  \label{fig:LC}
\end{figure}

In the SdS$_{w}$ spacetime, the LOE is given by \cite{HZ2017}  
\beqa
\label{eq:SdSwlightLC}
\ba{rcl}
\dis
\(\!\frac{1}{r^2}\frac{\d r}{\d \phi}\!\)^{\!\!2}
=\, \frac{\,1\,}{\,b^2\,}-\frac{1}{\,r^2\,}\!
\left[1\!-2\frac{M}{r}-2\!\left(\! \frac{\,\ro\,}{r}\!\right)^{\!3w+1}\right].
\ea
\eeqa
In figure~\ref{fig:LC}, we illustrate a trajectory of light. $\cal{N}$ denotes the point of closest approach of the trajectory, with $\(r,\phi\)=\(r_{*},\phi_{*}\)$. At this point, we have $\frac{\d r}{\d \phi}\big|_{\cal{N}}=0$; that is, 
\beqa
\label{eq:Npoint}
\ba{rcl}
\dis
 \frac{\,1\,}{\,b^2\,}=\frac{1}{\,r_{*}^2\,}\!
\left[1\!-2\frac{M}{r_{*}}-2\!\left(\! \frac{\,\ro\,}{r_{*}}\!\right)^{\!3w+1}\right].
\ea
\eeqa
The light orbit intersects with the circle of radius $r$ at two points, namely $\cal{P}$ and $\cal{Q}$. The Euclidean intersection angles $\beta_{E,\cal{P}}$ and $\beta_{E,\cal{Q}}$ are respectively given by
\beqa
\label{eq:betaE}
\ba{rcl}
\dis
\rm{tan}\,\beta_{\rm{E},\cal{P}}&=&\left[\frac{\it{r}\,\(\frac{\d \phi}{\d \lambda}\)}{\(\frac{\d \it{r}}{\d \lambda}\)}\right]_{\cal{P}}=\left[\it{r}\(\frac{\d \phi}{\d \rm{r}}\)\right]_{\cal{P}},\\[4mm]
\rm{tan}\,\beta_{\rm{E},\cal{Q}}&=&\left[\frac{\it{r}\,\(\frac{\d \phi}{\d \lambda}\)}{\(\frac{\d \it{r}}{\d \lambda}\)}\right]_{\cal{Q}}=\left[\it{r}\(\frac{\d \phi}{\d \rm{r}}\)\right]_{\cal{Q}}.
\ea
\eeqa
Since $\cal{P}$ and $\cal{Q}$ lie at the same radius, we have $\(\it{r}\,\frac{\d \phi}{\d \it{r}}\)_{\cal{P}}=-\(\it{r}\,\frac{\d \phi}{\d \it{r}}\)_{\cal{Q}}$ from the LOE \eqref{eq:SdSwlightLC}. Thus, one has
\beqa
\label{eq:P=Q}
\ba{rcl}
\dis
\beta_{\rm{E},\cal{P}}=-\beta_{\rm{E},\cal{Q}}.
\hspace*{10mm}
\ea
\eeqa
In fact, for any $r$, we can obtain the same result. 
In special, at the point of closest approach where $\cal{P}$ meets $\cal{Q}$, we have $\beta_{\rm{E},\cal{P}}=-\beta_{\rm{E},\cal{Q}}=-\pi/2$. 
It is now clear that $\left|\beta_{\rm{E},\cal{P}}\right|=\left|\beta_{\rm{E},\cal{Q}}\right|$ for any two points, $\cal{P}$ and $\cal{Q}$, at the same radius.
Combining these results, we finally conclude that the light orbit is symmetric relative to the straight line $\,\overline{\rm{O}\cal{N}}\,$ in the SdS$_{w}$ spacetime.

\vspace*{3mm}
\section{\hspace*{+0.0mm}Coordinate transformations}	
\label{app:C}
\vspace*{1mm}

Under the coordinate transformations
$\,r=\Omega\,\overline{r}$\, and $\frac{\d t }{\d\overline{\tau}}=\frac{\Omega}{\sqrt{1-2\left(\! \frac{\,\ro\,}{r} \!\right)^{\!3w+1}}}$,\,
we can rewrite the SdS$_{w}$ metric \eqref{eq:dSS2} as follows,
\vspace{0mm}
\beqa
\label{eq:OmegadS2}
\nonumber
\d S_{w}^2 = &-&\big[1\!-2\(\frac{\,M\,}{\Omega}\)\frac{1}{\bar{\it{r}}}-\!2\left(\! \frac{\,\ro\,}{\Omega} \!\right)^{\!3w+1}\left(\! \frac{\,1\,}{\bar{\it{r}}} \!\right)^{\!3w+1}\big]\, \frac{~~\Omega^2~~}{\big[1-\!2\left(\! \frac{\,\ro\,}{\Omega} \!\right)^{\!3w+1}\left(\! \frac{\,1\,}{\bar{\it{r}}} \!\right)^{\!3w+1}\big]}\d \overline{\tau}^{2}\\[4mm]
\nonumber
&+&\!\frac{(\frac{\d \Omega}{\d \bar{\it{r}}}+\Omega)^{2}}{\, \big[1\!-2\(\frac{\,M\,}{\Omega}\)\frac{1}{\bar{\it{r}}}-\!2\left(\! \frac{\,\ro\,}{\Omega} \!\right)^{\!3w+1}\left(\! \frac{\,1\,}{\bar{\it{r}}} \!\right)^{\!3w+1}\big]\,}\d \overline{r}^2\\[6mm]
\nonumber
&+&\Omega^2 \bar{r}^{2}\left(\d \theta^2\!+\sin^{2}\!\theta\,\d\phi^2\right), 
\eeqa
where the exact form of $\Omega=\Omega\(\bar{\it{r}}\)$ is given in He \& Zhang (2017) \cite{HZ2017} 
\beqa
\label{eq:Rbar-w}
\ba{rcl}
\dis
\Omega\(\bar{\it{r}}\) &=&\frac{\,\ro\,}{\,\rb\,}\!\left[
\frac{1}{2}\sin^2\!\(\!2\arctan\!\frac{\,\,(\rb/\ro)^{\frac{|3w+1|}{2}}}{\sqrt{2}}\!\)\!
\right]^{\!\frac{1}{|3w+1|}}\\[6mm]
&\simeq& 1- \frac{~(\rb /\ro)^{|3w+1|}\,}{|3w+1|}\simeq 1, ~~~~~~ \text{for}~~\rb^2 \ll r_o^2\,.
\ea
\eeqa
In the \textit{outer region} with $r\gg\rc$, $r$ is a monotonically increasing function of $\bar{\it{r}}$, since $\frac{\d r}{\d \bar{\it{r}}}\simeq\!1-\frac{3w}{3w+1}\!\(\! \frac{\,\ro\,}{\bar{\it{r}}}\!\)^{\!3w+1}>0$. 
Accordingly, we have $\bar{\it{r}}\,\simeq\,r$ from equation\,\eqref{eq:Rbar-w}. 
In this region, the Newtonian term can be ignored so that the SdS$_{w}$ metric \eqref{eq:dSS2} reduces to be conformally flat \cite{HZ2017}, as shown by equation\,\eqref{eq:pi/2dS^2}.
Therefore the light ray travels approximately along a physically straight line. Roughly, we can regard the outer region as a laboratory area.

\vspace*{3mm}
\section{\hspace*{+0.0mm}Calculations for the usual deflection angle}	
\label{app:D}
\vspace*{1mm}

Generally speaking, the traditional approaches for calculating the \textit{usual deflection angle} are based on integrating the LOE \cite{Ishak-Rev} 
rather than what we have done for the Gaussian deflection angle.
For comparison, we would like to calculate the usual deflection angle by using the new techniques presented in this work.
In gravitational lensing, the observer and source are usually located in the outer region with $r\gg\rc$, 
which can be approximately thought of as a laboratory area. 
Hereafter, we ideally assume that the SdS$_{w}$ metric reduces to be conformally flat in the outer region so that the contribution from this region to the deflection of light can be ignored. 
With a conformal transformation, the reduced metric can be written as $\rm{d}\it{S}^{{\rm 2}}={\rm \Omega}^{{\rm 2}}(\bar{\it{r}}) (- \rm{d}\bar{\tau}^{2}+\rm{d}\bar{\it{r}}^{2}+\bar{\it{r}}^{2} \,\rm{d}\phi^{2})$. 
In addition, the light ray needs to pass through the matter-dominated region, that is, $b\lesssim\rc$. Otherwise, if $b\gg\rc$, the light ray will travel along a physically straight line, just as the one traveling in the Minkowski spacetime;
we can not detect the bending of light, and are therefore unable to exact the information about the influence of dark energy on the bending of light.

As illustrated in figure~\ref{fig:LC}, we locate the source at point $\(\bar{\it{r}}_{\cal S},\phi_{\cal S}\)$ and the observer at point $\(\bar{\it{r}}_{\cal O},\phi_{\cal O}\)$, with $\rc\ll\bar{\it{r}}_{\cal O}, \bar{\it{r}}_{\cal S}\ll\ro$. 
Here we use $\phi_{\cal S}$ and $\phi_{\cal O}$ to denote the polar angles of the source and observer, respectively.
Since the source and observer are both far away from the mass center $M$ sitting at origin $\rm{O}$,
the path of the incident ray from the source and that of the outgoing ray arriving at the position of the observer can be described well by two straight lines in the outer region, 
respectively (see appendix~\ref{app:A} for details). 
By the symmetry of the light orbit with respect to to the straight line $\overline{\rm{O}\cal{N}}$ (see appendix~\ref{app:B} for details), the incident and outgoing rays should have the same impact parameter $b$. 
Thus, the two lines can be described by 
\beqa
\label{eq:straightLines}
\ba{rcl}
\dis
\bar{\it{r}}=\frac{b}{\rm{cos}\,(\phi\!-\!\phi_{\rm{i}})}\,\,\,\,\, \rm{and}\,\,\,\,\bar{\it{r}}=\frac{{\it b}}{\rm{cos}\,(\phi\!-\!\phi_{\rm{f}})},
\ea
\eeqa
with $\phi_{\rm{i}}$ and $\phi_{\rm{f}}$ being the their polar angles from the positive $x$-axis, respectively. 
We define the two angles $\phi_{\rm{i}}$ and $\phi_{\rm{f}}$ concretely, as in depicted figure~\ref{fig:LC}. 
In particular, we have $\phi_{\rm{i}}\simeq\phi_{\cal S}$ and $\phi_{\rm{f}}\simeq\phi_{\cal O}$ in the outer region $r\,\gg\,\rc\,\gtrsim\,b$.
Denote $K_{\rm{in}}^{\,a}$ as the four-momentum vector of the incident light ray and $K_{\rm{ou}}^{\,a}$ as that of the outgoing one. 
They are both radial vectors so that their angular components $K_{\rm{in}}^{\phi}$ and $K_{\rm{ou}}^{\phi}$ are both zero: $K_{\rm{in}\perp}^{\,\,\,a}=(\Omega\,K_{\rm{in}}^{\bar{\it{r}}})\, \textit{e}_{\bar{\it{r}}}^{\,a}$ with $\phi=\psi_{\rm{i}}$, and $K_{\rm{ou}\,\perp}^{\,\,\,a}=(\Omega\,K_{\rm{ou}}^{\bar{\it{r}}}) \,\textit{e}_{\bar{\it{r}}}^{\,a}$ with $\phi=\psi_{\rm{f}}$. 
Here, $\,\psi_{\rm{i}}\,$ and $\,\psi_{\rm{f}}\,$ are correspondingly the polar angles of these two vectors, respectively.
In physics, they are just the incident and outgoing angles, respectively.
Combining the definitions of $\phi_{\rm{i}}$ and $\phi_{\rm{f}}$, we have $\psi_{\rm{i}}=\phi_{\rm{i}}+\pi$ and $\psi_{\rm{f}}=\phi_{\rm{f}}$, as shown in figure~\ref{fig:LC}. 
In fact, both $\psi_{\rm{i}}$ and $\psi_{\rm{f}}$ are physically measurable angles; exactly, they are both measured from the $x$-axis that is actually the reference null geodesic with $b=0$.

In figure~\ref{fig:LC}, the light ray travels along the path $L\doteq\cal{S}\!\!\to\!\!_{\cal{P}}\!\!\to\!\!_{\cal{N}}\!\!\to\!\!_{\cal{Q}}\!\!\to\!\!\cal{O}$ passing through the matter-dominated region with $\,r\lesssim\rc$. 
Recalling the parametrization for the $\gamma$ curve shown in section~\ref{sec:4}, we take the parallel transport of $V^{a}\(0\)=K_{\rm{in}}^{\,a}$ along the path $L$, and then get the corresponding vector $\bar{V}^{a}(\bar{\lambda}_{0})=K_{\rm{ou}}^{\,a}$ with $\bar{\varphi}\(\bar{\lambda}_{0}\)=\psi_{\rm{f}}$ at point $\cal{O}$. On the other hand, we perform the parallel transport of $V^{a}\(0\)=K_{\rm{in}}^{\,a}$ along $L_{0}$ in the outer area where $r\simeq\bar{\it{r}}\,\gg\,\rc$ (see appendix~\ref{app:C} for details), and obtain $V^{a}\(\lambda_{0}\)=K_{\rm{in}}^{\,a}$ with $\varphi\(\lambda_{0}\)=\varphi\(0\)$ at point $\cal{O}$. 
Clearly, we have $\varphi\(0\)=\psi_{\rm{i}}\(=\phi_{\rm{i}}+\pi\)$.
We therefore obtain the {\it usual deflection angle} $\alpha_{M}$ between the incident and outgoing rays: $\alpha_{M}=\bar{\varphi}\(\bar{\lambda}_{0}\)-\varphi\(0\)=\phi_{\rm{f}}-\phi_{\rm{i}}-\pi$, which is also physically measurable.
With $u=\rm{1}/r$, integrating the LOE \eqref{eq:SdSwlightLC} yields the following formula,
\beqa
\label{eq:SdSwlightLCU}
\ba{rcl}
\dis
\alpha_{M}\simeq\phi_{\cal O}-\phi_{\cal S}-\pi=\sum\nolimits_{\,p\,=\,\tiny{\cal{S}},\,\tiny{\cal{O}}}\mathlarger{\int}^{u_{*}}_{u_{p}}\!\!\!\frac{\d u}{\sqrt{1/b^{2}\!-\!u^{\rm{2}}\!+\!\rm{2} \it{M}\it{u}^{\rm{3}}\!+\!\rm{2}\it{r}_{o}^{\rm{3}\it{w}\!+\!\rm{1}}\it{u}^{\rm{3}\(\!\it{w}\!+\!\rm{1}\!\)}}}-\pi,~~~~~~
\ea
\eeqa
where $u_{p=\cal{S}}=1/r_{\cal{S}}$, $u_{p=\cal{O}}=1/r_{\cal{O}}$, and $u_{*}=1/r_{*}$. Here, $r_{\cal{S}}$ and $r_{\cal{O}}$ are the radii of the source and observer in the original coordinates $\(r, \phi\)$, respectively. 
For the special case of the cosmological constant ($\,w=-1$),  by taking $u_{\cal{O}}\to\!0^{+}$ and $u_{\cal{S}}\to\!0^{+}$, we have $\alpha_{M}\simeq\,\frac{4M}{b} [1+(\frac{b}{r_{o}})^2]$. 
When setting $\CC=0$, we recover the conventional result of light bending, $\alpha_{M}\simeq\,\frac{4M}{b}$. 
For $\,w\neq-1$, it is difficult to approximate the formula \eqref{eq:SdSwlightLCU} analytically in a simple form. 
Anyway, after considering the measurements made by the static observer and adopting new techniques, we provide a conceptually clean and independent resolution by using a traditional approach to the issue of light deflection in the SdS$_{w}$ spacetime.

However, the approach used here is not based on a strict mathematical basis. 
For instance, we can not take $r\to\infty$, since otherwise it will go beyond the outer horizon $r\sim\ro$. 
Thus, compared with the $w$-term, the newtonian term in the SdS$_{w}$ metric can not be neglected completely. 
Hence, the conformal-flatness of the outer region with $r\gg\rc$ is not as exact as we assumed. 
Besides, even though the effect of dark energy on the bending of light can be locally neglected in the outer region, it may be amplified significantly after the travel of the light ray over a long distance.
So we have to rethink the traditional approaches.
There is still no any strict way to deal with the outer region. 
It is the non-conformal flatness of the outer region that has led a long-term debate on the topic \cite{Ishak2008,Arakida12,Italy,Sereno08,Bhadra10,other1,Ishak2010,Ishak-Rev,Gibbons08,Ishihara16,Ishihara17,Arakida18} 
In fact, it is almost impossible to define the usual deflection angle strictly in the SdS$_{w}$ spacetime.
Now, by comparisons, it can be clear that, via equations~\eqref{eq:GaussianCurvature} and \eqref{eq:polygon}, 
we provide a new way to solve the problem on whether dark energy affects the bending of light, 
which completely avoids the non-conformal-flatness problem that we have to encounter in the traditional approaches.

\vspace*{3mm}
\section{\hspace*{+0.0mm}The Gauss-Bonnet theorem and its application}	
\label{app:E}
\vspace*{1mm}

For any global surface or two-dimensional Riemannian manifold $\Sigma$, it can be described locally by the metric \cite{Carmo16,Chern00} 
\beqa
\label{eq:ds2_2d}
\ba{rcl}
\dis
\d s^{2}&=&g_{ij}\,\rm{d}\it{z}^{i}\rm{d}\it{z}^{j}\\[2mm]
&=&\it{E}\,\rm{d}\it{u}^{\rm{2}}+\rm{2}\it{F}\,\rm{d}\it{u}\,\rm{d}\it{\upsilon}+\it{G}\,\rm{d}\it{\upsilon}^{\rm{2}},
\ea
\eeqa
where $\Sigma$ is parametrized by the local coordinates $\(\it{z}^{i}, \it{z}^{j}\)=\(u, \upsilon\)$.
Denote $\d \sigma$ as the element of area over $\Sigma$, and $K$ as the Gaussian Curvature. 
In general, $\d \sigma=\sqrt{EG-F^{2}}\,\rm{d}\it{u}\,\rm{d}\it{\upsilon}$ \cite{Carmo16,Chern00}. 
In GR, the surface $\Sigma$ can be globally described by a metric; in fact, the entire spacetime is uniquely determined by one metric.
By the definition of the Gaussian Curvature $K$, it can be expressed as  \cite{Rindler06,Chern00} 
\beqa
\label{eq:GaussK0}
\ba{rcl}
\dis
K=\frac{R_{u\upsilon\,\!u\upsilon}}{g_{uu}\,g_{\upsilon\upsilon}-g_{u\upsilon}\,g_{u\upsilon}}, 
\ea
\eeqa
where $R_{u\upsilon\,\!u\upsilon}$ denotes a specific component of the Riemann tensor of rank 4.

Denoted by $D \(\subseteq\Sigma\)$ the simple, connected region bounded by the closed curve $\gamma$. 
Assume that $\gamma$ is positively oriented, parametrized by arc length $s$, and let $\theta_{i}$ and $\gamma\(s_{i}\)$ be, respectively,  the $i$th external angle and the $i$th vertex of $\gamma$. Then \cite{Carmo16,Chern00}
\beqa
\label{eq:GBT2}
\ba{rcl}
\dis
\sum_{i}\int^{s_{i+1}}_{s_{i}}k_{g}\d s+\int\!\!\!\int_{D}\,K\d \sigma+\sum_{i}\theta_{i}=2\pi,
\hspace*{0mm}
\ea
\eeqa
where $k_{g}=k_{g}\(s\)$ is the geodesic curvature of the regular arcs of $\gamma$. 
This is the famous Gauss-Bonnet theorem in global differential geometry. It establishes a connection between local and global properties of curves and surfaces.
Following this theorem, we can derive the following formula
\beqa
\label{eq:GBtoPT2}
\ba{rcl}
\dis
\Delta\varphi=\varphi\(l\)-\varphi\(0\)=\int\!\!\!\int_{D}\,K\d \sigma, 
\hspace*{0mm}
\ea
\eeqa
which is independent of the choice of coordinates \cite{Carmo16,Chern00}.

In fact, the metric \eqref{eq:ds2_2d} can be chosen to be orthogonal by coordinate transformations \cite{Carmo16}. 
Then $F=0$. $K$ therefore can be simply expressed as 
\beqa
\label{eq:GaussK}
\ba{rcl}
\dis
K=-\frac{1}{2\sqrt{EG}}\(\(\frac{(\sqrt{E})_{\!\upsilon}}{\sqrt{G}}\)_{\!\!\!\upsilon} +\(\frac{(\sqrt{G})_{\!u}}{\sqrt{E}}\)_{\!\!\!u}\), 
\ea
\eeqa
where $f_{u}$ ($f_{\upsilon}$) denotes the partial derivative of $f$ with respect to $u$ ($\upsilon$). 
It follows that the total curvature $K_{\rm{tot}}$ can be described by the following form, 
\beqa
\label{eq:total}
\nonumber
K_{\rm{tot}}&=&\int\!\!\!\int_{D}\,K\d \sigma\\
\nonumber
&=& -\frac{1}{2}\int\!\!\!\int_{D}\,\(\(\frac{(\sqrt{E})_{\!\upsilon}}{\sqrt{G}}\)_{\!\!\!\upsilon} +\(\frac{(\sqrt{G})_{\!u}}{\sqrt{E}}\)_{\!\!\!u}\)\,\rm{d}\it{u}\,\rm{d}\it{\upsilon}\\
&=&-\frac{1}{2}\oint_{\partial{\rm D}}\(-\frac{(\sqrt{E})_{\!\upsilon}}{\sqrt{G}}\,\rm{d}\it{u}+\frac{(\sqrt{G})_{\!u}}{\sqrt{E}}\,\rm{d}\it{\upsilon}\),
\hspace*{0mm}
\eeqa
where $\partial{\rm D}$ is the boundary of $D$. The first line is simply the definition of the total curvature. 
The second line comes directly from equation\,\eqref{eq:GaussK}.
The last line uses Green’s theorem.


\end{document}